\documentclass[sigconf]{acmart}
\DeclareMathSizes{10}{11}{7}{5}

\setcopyright{none}

\usepackage[english]{babel}
\usepackage{blindtext}

\usepackage{tabularx,tablefootnote,wrapfig,multirow}
\usepackage{hyphenat,xspace,color,colortbl,enumitem}
\usepackage{booktabs,multirow,microtype}
\usepackage{bm}
\usepackage[normalem]{ulem}
\usepackage{newtxmath,bm,courier,textcomp}
\usepackage{enumitem,fancyref}
\usepackage{xfrac,sparklines,bigstrut,physics}
\usepackage[labelformat=simple,skip=0pt]{subcaption}
\usepackage{amsmath,amsfonts,amsbsy}
\usepackage[scaled=0.90]{helvet}

\newcommand{\systemname}{ParaMax}
\newcommand{\systemnames}{ParaMax's}
\def\argmin#1{\underset{#1}{\operatorname{arg\,min}}}

\definecolor{Gray}{gray}{0.9}
\definecolor{LightGreen}{rgb}{0.88,1,0.88}
\definecolor{DarkGreen}{rgb}{0.0,0.4,0.13}
\definecolor{LightOrange}{rgb}{1,0.85,0.8}
\definecolor{LightYellow}{rgb}{1,1.00,0.5}
\definecolor{LightRed}{rgb}{1,0.80,0.80}

\usepackage{soul}

\newcommand{\RN}[1]{%
  \textup{\uppercase\expandafter{\romannumeral#1}}%
}

\newcommand{\parahead}[1]{\vspace{2pt plus 0pt minus 2pt}\noindent{\bfseries #1}}
\newcommand{\parabreak}{\vspace*{1.00ex minus 0.25ex}\noindent}

\renewcommand{\paragraph}[1]{\parahead{#1}}

\usepackage{fontawesome}
\usepackage{tikz}
\usepackage{amsmath}

\usepackage{efbox,graphicx}
\efboxsetup{linecolor=blue,linewidth=0.8pt}

\copyrightyear{2021}
\acmYear{2021}
\setcopyright{usgovmixed}\acmConference[ACM MobiCom '21]{The 27th Annual
International Conference on Mobile Computing and Networking}{October 25--29,
2021}{New Orleans, LA, USA}
\acmBooktitle{The 27th Annual International Conference on Mobile Computing and
Networking (ACM MobiCom '21), October 25--29, 2021, New Orleans, LA, USA}
\acmPrice{15.00}
\acmDOI{10.1145/3447993.3448619}
\acmISBN{978-1-4503-8342-4/21/10}

\begin{document}
\fancyfoot[L,R,C]{} 


\title{Physics-Inspired Heuristics for Soft MIMO Detection\\ in 5G New Radio and Beyond\vspace{-0.5cm}}




\author{Minsung Kim$^{\star,\dagger,\ast}$, Salvatore Mandr\`a$^{\dagger,\top}$, Davide Venturelli$^{\ast}$, Kyle Jamieson$^{\star}$}


\affiliation{%
 \institution{\small
 Princeton University$^{\star}$, NASA Ames Research Center, QuAIL$^{\dagger}$, USRA Research Institute for Advanced Computer Science$^{\ast}$, KBR, Inc.$^{\top}$}}

\begin{abstract}
Overcoming the conventional trade-off between throughput and 
bit error rate (BER) performance, versus
computational complexity is a long-term challenge for uplink Multiple-Input
Multiple-Output (MIMO) detection in 
base station design for the cellular 5G New Radio roadmap, as well as in
next generation wireless local area networks. In this work, we 
present \textbf{\systemname{}}, a  MIMO detector architecture that
for the first time brings to bear 
physics-inspired parallel tempering algorithmic 
techniques \cite{swendsen1986replica, metropolis1949monte, hastings1970monte}
on this class of problems.
\systemname{} can achieve near optimal maximum\hyp{}likelihood (ML)
throughput performance in the Large MIMO regime,
Massive MIMO systems where the base station
has additional RF chains, to approach the number
of base station antennas, in order to support even more parallel
spatial streams.
\systemname{} is able to achieve a near ML-BER performance up to $160\times 160$ and $80\times 80$ Large MIMO for low-order modulations such as BPSK and QPSK, respectively, only requiring less than tens of processing elements. 
With respect to Massive MIMO systems, in $12\times 24$ MIMO with 16-QAM at SNR 16~dB, \systemname{} achieves 330~Mbits/s near-optimal system throughput with 4\hyp{}8 processing elements per subcarrier, which is  approximately $1.4${\small $\times$} throughput than linear detector-based Massive MIMO systems.



\end{abstract}



\settopmatter{printacmref=true, printccs=true, printfolios=true}


\maketitle{}

\thispagestyle{empty} 

\section{Introduction}
\label{s:intro}

\begin{figure}
\centering
\includegraphics[width=0.87\linewidth]{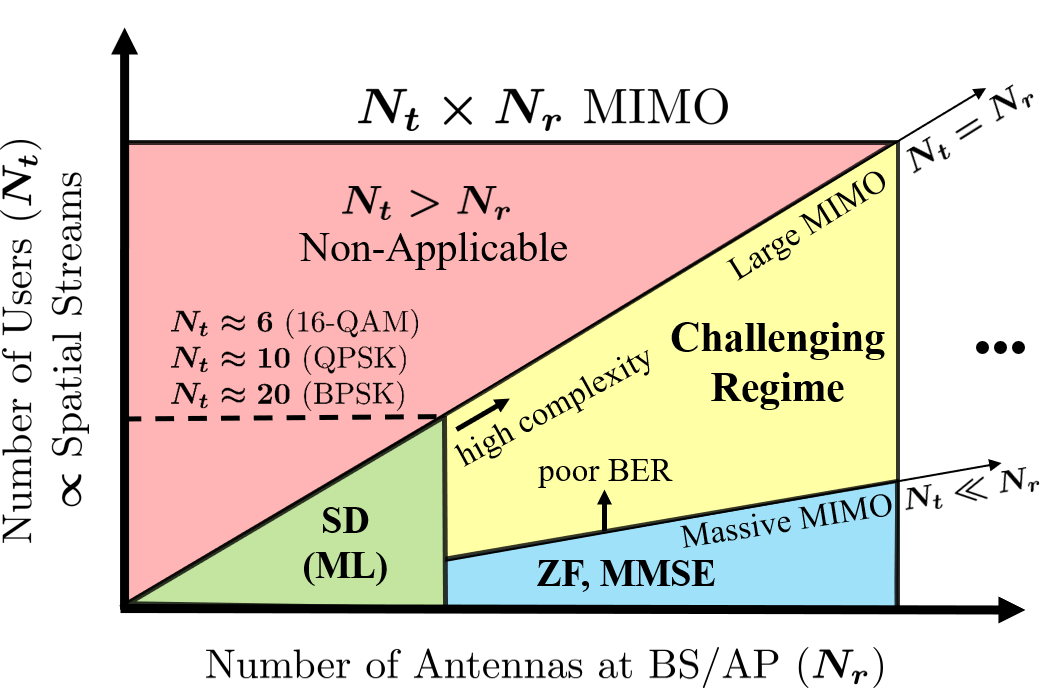}
\caption{\normalfont Fundamental MIMO regimes in 5G New Radio and next 
generation local-area networks, and approximate feasibility of 
various detection approaches.}
\label{f:mimo_regime}
\end{figure}
Multi-User Multiple-Input Multiple-Output (MU-MIMO) has proven an
essential technique to maximize capacity in many different kinds of wireless systems 
such as 802.11 wireless LAN and 5G New Radio cellular networks. In MU-MIMO, the
uplink receiver (\emph{i.e.}, an \emph{access point}---AP---in a wireless 
LAN, or a \emph{base station}---BS---in a cellular network) with 
multiple antennas supports many users simultaneously by striping data over parallel streams 
(a technique known as \emph{spatial multiplexing}), and thus enables significantly
higher data capacities.  In an ideal world, the number of parallel streams that MU-MIMO
can support would be the lesser of the number of mobile users and the number of 
radios at the base station, and overall system capacity 
would increase proportionally to the number of spatial streams.

In practice, however, the \emph{channel hardening} phenomenon
complicates this situation, in the following way.  
MU-MIMO requires signal processing to disentangle the spatial
streams from each other, a technique called \emph{MIMO detection}. 
For a base station
with as many antennas as radio front ends, when the number
of users approaches the number of base station antennas, 
MIMO detection becomes extremely difficult
resulting in poor performance for conventional linear detection algorithms \cite{BigStation}: this is the
\emph{Large MIMO} regime that lies along
the points where the number of users $N_t$ equals
the number of base station antennas $N_r$, as depicted 
in Figure~\ref{f:mimo_regime}.\footnote{For simplicity, we call $N_t \times N_r$ MIMO regimes, ``Large MIMO'' when $N_t=N_r$, while ``Massive MIMO'' when $N_t<N_r$, regardless of $N_t$ size.}
For Large MIMO, there exist \emph{maximum-likelihood} (ML) \emph{exact}
solvers, that can achieve the lowest possible bit error rate and,
therefore, restore a high throughput. Unfortunately, these detection algorithms come at the expense of 
an exponential increase of the required computational resources 
as MIMO size increases, eventually becoming infeasible for many users
because of the processing time limits in wireless systems.  
For example, at most 
three milliseconds of BS's computation are available for both the 4G
LTE uplink and downlink~\cite{BigStation, dahlman20134g}. 

\emph{Massive MIMO} systems such as LuMaMi and Lund (6-12 users, 100-128 BS antennas)
\cite{rusek2012scaling, vieira2014flexible,malkowsky2017world}, Argos 
(eight users, 96 BS antennas) \cite{argos-mobicom12, 
argos-asilomar17}, BigStation \cite{BigStation}, Agora~\cite{ding2020agora}, and Samsung's
5G base stations (16 users, 64 BS antennas) \cite{samsung}  
mitigate channel hardening 
in the following way. Since linear detectors such as \emph{Zero\hyp{}Forcing} 
(ZF) and \emph{Minimum Mean\hyp{}Squared 
Error} (MMSE) can achieve near\hyp{}ML
performance when the wireless channel is well\hyp{}conditioned,
systems that use many more base station antennas than users\fshyp{}spatial streams
(\emph{i.e.,} $N_r \gg N_t$ for $N_t\times N_r$ MIMO) may offer each base station radio a choice
of one out of a number of antennas to use.  This largely negates the effect of
channel hardening, but requires base station antennas numbering 
a sufficient factor greater than users (\emph{e.g.} $N_r \ge 10N_t$~\cite{bjornson2015massive} or $N_r \ge 4N_t$ for 16 users or below~\cite{argos-asilomar17}, while there is no proven rule-of-thumb of $N_r/N_t$ that maximizes the spectral efficiency~\cite{bjornson2015massive}),
as shown in Figure~\ref{f:mimo_regime}, to achieve the full 
throughput of $N_t$ spatial streams.
In addition, the deployment of larger numbers of antennas 
eventually becomes challenging from a practical standpoint, most acutely
in wireless local area networks, but also in small, densely\hyp{}deployed
5G base stations where form factors preclude excessive numbers of antennas,
and eventually in normal base stations where tower size faces practically 
limited.

In this paper, we take a complementary approach to Massive MIMO: we begin
with a particular Massive MIMO configuration in which the number of 
base station antennas is practically at its maximum, and then ask the 
question \emph{how can performance be further improved via 
additional spatial streams?}  The answer lies in a fusion of two
preceding ideas: add radio chains at the Massive MIMO
base station to equal the number of antennas, and at the same
time, utilize near\hyp{}ML detection algorithms.  This pushes us
out towards the upper\hyp{}right corner of the space in Figure~\ref{f:mimo_regime}
and maximizes computational complexity, yet offers the promise of the greatest
spatial multiplexing gains, given our practical constraint on
base station antenna count.

\parahead{A shift to Physics-inspired approaches.}
Over the last few years, 
there has been a surge of interest in alternative computation approaches to reduce
the complexity of current detectors by leveraging algorithms that relate 
optimization convergence to Physics principles. This interest 
is further accelerating in view of experimental initiatives featuring hardware-native 
implementations of these approaches, using both quantum and classical physics-based computations~\cite{kim2019leveraging, hamerly2019, aramon2019physics, csaba2020coupled, Hadfield:2017:QAO:3149526.3149530,kasi2020towards,kim2020towards}.
One common aspect of these algorithms is that they frame the 
computational problem as an energy minimization problem of a magnetic spin system,
also known as the \emph{Ising spin model}~\cite{ising1925beitrag}.
Beside being an important model to understand the physics behind magnetic 
systems, \emph{any} NP computational problem can be expressed as the energy 
minimization of an appropriate Ising spin model \cite{lucas2014ising} (that is, the Ising spin model 
is NP-Complete \cite{welsh1990computational}).
In this regard, physics\hyp{}inspired algorithms
can be seen as parametric ``black boxes'' that accept an Ising spin problem as input,
and output the configuration with lowest associated energy. What distinguishes one algorithm from another is the underlying mechanism used to find the global minimum, which corresponds to the ML optimal solution in MIMO detection.



\parabreak{}This paper presents the design and implementation 
of \textbf{\systemname{}}, a soft MU-MIMO detector system for Large and Massive MIMO
networks that uses \emph{parallel tempering}, a 
physics-inspired heuristic algorithm, on classical platforms. 
\systemname{} operates flexibly in parallel for any number of available processors, 
supporting fixed latency and highly\hyp{}scalable parallelism. 
We design the \emph{ParaMax Ising Solver} (\S\ref{s:PMIS}),
a parallel tempering\hyp{}based
solver that is tailored for MIMO detection, implemented as
a fully classical algorithm that does not require any specific 
hardware, and integrate it into the overall design of
our system (\S\ref{s:system_archtitecture}).
We also introduce a new algorithm (\S\ref{s:detection_confidence})
to generate soft information for 
heuristic detectors that enables a more reliable detection and 
decoding. The proposed algorithm utilizes heuristic detection outputs and generates soft information,
defined as the
bitwise detection confidences that implicitly take channel conditions 
and noise into consideration. 
To our best knowledge, this is the first application of parallel tempering 
to wireless networks, and \systemname{} is the first heuristic-based 
MIMO detector that demonstrates near\hyp{}ML performance for both very Large 
and Massive MIMO successfully. 

\begin{figure}
\centering
\includegraphics[width=0.9\linewidth]{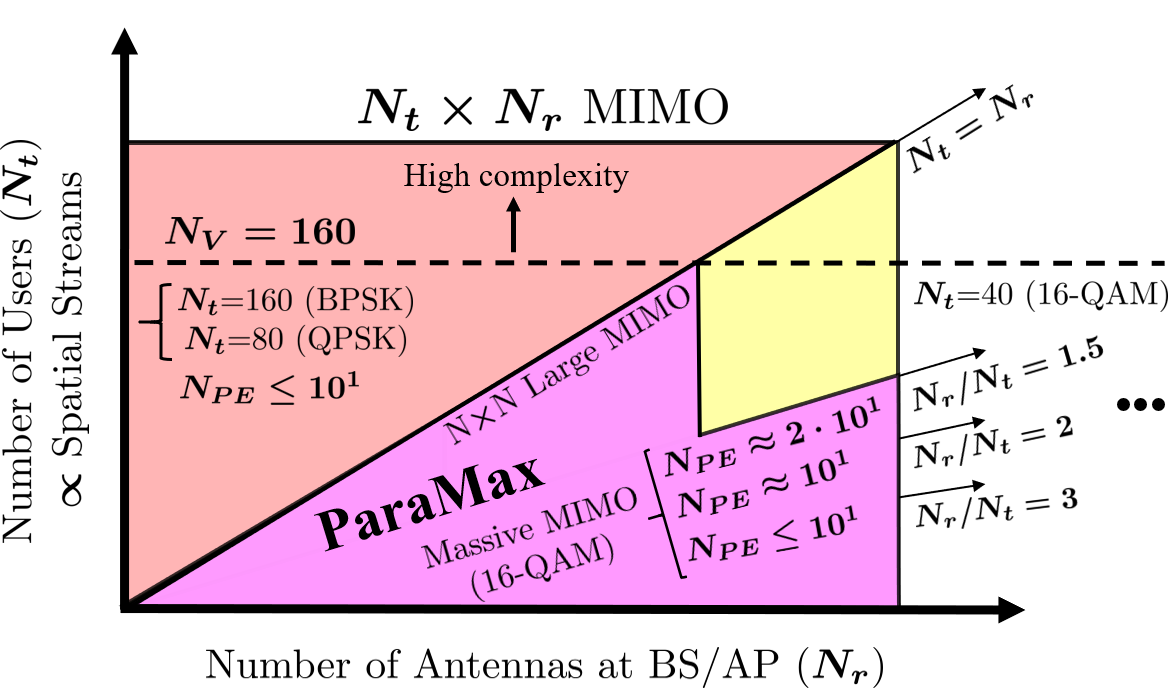}
\caption{\normalfont Summary of \systemnames{} feasible MIMO regimes
(\emph{cf.}~Figure~\ref{f:mimo_regime}) and required 
processing element count $N_{\mathrm{PE}}$ per subcarrier.}
\label{f:summary}
\end{figure}

Our experiments show that \systemname{} achieves a 
constantly\hyp{}increasing performance as the number of processing elements 
increases. 
In the case of lower\hyp{}order
BPSK and QPSK modulations, very large MIMO of $160 \times 160$ and $80 \times 80$ respectively, can achieve near-ML performance
for less than tens of processing elements, 
as depicted in Figure~\ref{f:summary}. With respect to Massive MIMO systems,
in $12\times 24$ MIMO with 16-QAM modulation at SNR 16~dB, \systemname{} 
achieves a 330~Mbits/s near-optimal system throughput with 4\hyp{}8 processing elements (PEs) per subcarrier, 
approximately $1.4${\small $\times$} better
throughput than linear detector\hyp{}based Massive MIMO systems.

\section{Background}
\label{s:primer}

This section introduces background knowledge, indicating relevant literature. 
Sections~\ref{s:primer_sa} and~\ref{s:primer_pt} 
respectively explain \systemnames{} algorithmic foundations, simulated 
annealing and parallel tempering. Section~\ref{s:primer_mimo_detection} 
describes the MU-MIMO model and detection problem.

\vspace{-0.2cm}
\subsection{Simulated Annealing}
\label{s:primer_sa}

\emph{Simulated annealing} (SA) is a classical heuristic optimization technique typically used to find the state or \emph{configuration} $\mathbf{s}$ with the lowest energy of \emph{Ising spin} problems, where $\mathbf{s}$ is a vector consisting of
$\{s_1,s_2,\cdots,s_{N_V}\}$ spins, with each spins $s_i$ assuming the
values \{-1,\,+1\}.
%
%
In general, the energy objective function of Ising spin problems (also called \emph{Hamiltonian}) is represented as a quadratic cost function of the following form:
\begin{equation}\label{eq:ising-ham}
\mathcal{H}(\mathbf{s}) = \sum_{ij} g_{ij} s_i s_j + \sum_i f_i s_i,
\end{equation}
with $g_{ij}\in\mathbb{R}$ being \emph{(anti-)ferromagnetic couplings} that indicate a preference of correlation ($s_i=s_j$ or $s_i\neq s_j$) between two spins, and $f_i \in\mathbb{R}$ \emph{local magnetic fields} that individually
act on $s_i= \pm1$. Any optimization problem, including 
MIMO detection, can in theory be translated to an 
Ising spin model by properly choosing the $\{g_{ij}\}$ and $\{f_i\}$
\cite{welsh1990computational, lucas2014ising}.

\parahead{Simulated Annealing (SA) Heuristic.} 
SA is inspired by the physical process of \emph{annealing}, where a 
metallic material is slowly cooled from high temperature to 
eventually reach a molecular state or atomic configuration where 
the potential energy of the material is minimized. SA 
numerically emulates this process in order to 
find the global optimum (or \emph{ground state}) of 
Eq.~\ref{eq:ising-ham}. To enable SA, it is necessary
to simulate a \emph{thermal bath} which imitates the cooling or 
annealing process interacting with the Ising spin model. More 
precisely, the probability that a given spin\hyp{}configuration 
$\mathbf{s}$ is explored by the Ising spin system at a given
\emph{inverse temperature} $\beta = 1/T$ follows the 
\emph{Gibbs distribution} $p(\mathbf{s}) = 
\exp[-\beta\mathcal{H}(\mathbf{s})]/\mathcal{Z}$,
with $\mathcal{Z}$ usually called \emph{partition} function~\cite{georgii2011gibbs, crooks2007measuring}.
As the temperature $T$ is lowered, the probability 
$p(\mathbf{s})$ of finding a
state $\mathbf{s}$ with an energy larger than the minimum
energy becomes exponentially lower.
Therefore, sampling from the low-temperature 
Gibbs distribution allows rapid
detection of the spin configuration with the lowest energy
with high probability. 


However, the calculation of the partition function $\mathcal{Z}$, and thus $p(\mathbf{s})$, is computationally challenging, particularly for low temperatures.
To avoid the direct calculation of the Gibbs distribution $p(\mathbf{s})$, Metropolis \emph{et al}.~\cite{metropolis1949monte} proposed the use of 
Markov chain processes to help the system 
emulate the annealing and heuristic exploration of configurations at a given temperature. Specifically, they 
proposed a random process
to ``flip'' a spin, with probability depending only
on temperature and the Hamiltonian (but not on $\mathcal{Z}$), \emph{i.e.}:
\vspace{-0.15cm}
\begin{equation}\label{eq:metropolis}
    p(s_i \rightarrow -s_i) = \min\left\lbrace1, e^{-\beta\Delta\mathcal{H}}\right\rbrace,
\end{equation}
with $\Delta\mathcal{H}$ the variation of energy once 
the spin $s_i$ ($\forall i$) is flipped for a given initial configuration. 
Hence,
moves that would eventually reduce the overall energy of the spin system are always accepted. Otherwise,
there is a chance that such spin flip is either accepted or rejected. 
Metropolis \emph{et al}. showed that the spin system will eventually thermalize
to the corresponding temperature if the rejection rule in Eq.~\ref{eq:metropolis} 
(also called \emph{Metropolis updates} or \emph{sweeps})
is iteratively applied. Therefore, it is in principle possible to find the lowest energy spin configuration and, consequently,
the solution to the original problem, by starting from a very large temperature and slowly decreasing
it by iteratively applying the rule in Eq.~\ref{eq:metropolis}.
\vspace{-0.2cm}

\subsection{Parallel Tempering}
\label{s:primer_pt}

SA guarantees that the spin system
will eventually find the lowest energy spin configurations if the temperature is lowered
slowly enough. However, for hard optimization problems, it may require an exponentially long time. Indeed,
a rugged energy landscape ``traps'' the spin system in local minima which are hard to escape: \emph{parallel tempering}~\cite{swendsen1986replica} helps the spin system escaping local minima and, therefore, thermalize
faster at a low temperature. 
The basic principle of parallel tempering is simple: instead of a a single spin system, different \emph{replicas} are simulated in parallel, each with
a different temperature. After a certain amount of Metropolis updated, the temperatures of the
two replicas $r_1$ and $r_2$ are exchanged following the updating rule:
\begin{equation}\label{eq:pt}
    p(r_1 \leftrightarrow r_2) = \min\left\lbrace{1, e^{\Delta\beta\Delta\mathcal{H}}}\right\rbrace,
\end{equation}
with $\Delta\beta$ and $\Delta\mathcal{H}$ being the difference in the inverse of temperature 
and the difference in energies of the two replicas respectively. As one can see, two temperatures
are always exchanged if a replica at higher temperature has a lower energy than a replica
with a lower temperature. Otherwise, the exchange of the two temperatures is either accepted or
rejected accordingly to Eq.~\ref{eq:pt}. 
In a variety of hard optimization problems, parallel tempering drastically speeds up the thermalization of the spin system \cite{katzgraber2006feedback, young2004absence, trebst2006optimized}, 
including benchmark against quantum annealers \cite{mandra2016strengths, mandra2017exponentially, mandra2018deceptive}.
\vspace{-0.2cm}
\subsection{MIMO Detection}
\label{s:primer_mimo_detection}


The input and output relationship of a spatial multiplexing MIMO system (per subcarrier in OFDM systems~\cite{nee2000ofdm}) with $N_t$ input antennas at user side (or $N_t$ single-antenna users for simplicity) and $N_r$ output antennas ($N_t \leq N_r$) with $N_R$ radios ($N_R \leq N_r$) at the receiver side is described as $\mathbf{y} =  \mathbf{H\bar{v}}+\mathbf{n}$. With $N_t \leq N_R$ (\emph{i.e.,}  $N_t \times N_r$ MIMO with $N_t$ radio streams), here $\mathbf{y} \in{\mathbb{C}^{N_r}}$ is the received vector perturbed by \emph{additive white Gaussian noise} (AWGN) $\mathbf{n}$ $\in{\mathbb{C}^{N_r}}$, $\mathbf{H}$ $\in{\mathbb{C}^{N_r\times N_t}}$ is the wireless channel, and $\mathbf{\bar{v}} \in \mathcal{O}^{N_t}$ is the transmitted set of $N_t$ symbols with \emph{constellation} $\mathcal{O}$ (\emph{e.g.}, 4-, 16-, 64-QAM), representing $N_V =N_t\, \log_2{|\mathcal{O}|}$ bits per channel use, with $|\mathcal{O}|$ the size of the modulation. 
MIMO detection at the receiver side (AP or BS) is a technique to find a candidate solution $\mathbf{\hat{v}}\in \mathcal{O}^{N_t}$ with an objective of detecting the transmitted symbol vector (\emph{i.e.}, the objective is $\mathbf{\hat{v}=\bar{v}}$) based on the received signal $\mathbf{y}$ and estimated $\mathbf{H}$. Pilot symbols enable the estimation of $\mathbf{H}$.

\parahead{Maximum Likelihood Detection.}
\emph{Maximum likelihood detection} (ML detection) is optimal in the
sense that it minimizes the error probability of the detection. 
It is defined as
\begin{equation}   
\begin{small}
\label{eqn:ml}
\hat{\mathbf{v}} = \argmin{\mathbf{v} \in \mathcal{O}^{N_t}}
\left\lVert\mathbf{y} - \mathbf{Hv}\right\rVert^2.
\end{small}
\end{equation}
The search set of $\mathbf{v}$ (\emph{i.e.}, the \emph{search space} \mbox{$\mathbb{S}\subseteq O^{N_t}$}) is the set of possible solutions that the
optimizer can take into account. 
Each element $\mathbf{v}$ in the search space is a candidate solution 
with which the values of the \emph{ML objective function} 
$\mathcal{D}(\mathbf{v}) = \left\lVert\mathbf{y} - \mathbf{Hv}\right\rVert^2$ 
in Eq.~\ref{eqn:ml} (\emph{i.e.}, Euclidean distances) are measured and 
compared with each other. The best candidate, with minimum value, 
becomes the ML solution $\mathbf{\hat{v}}$. Further, $\mathbb{S}$ is 
an indicator of complexity.  
In principle, the ML search involves all possible candidates (\emph{i.e.}, $\mathbb{S}= \mathcal{O}^{N_t}$) 
which makes the brute-force approach intractable for large MIMO sizes with high-order modulations.  




\parahead{Sphere Decoder.} 
The \emph{Sphere Decoder} (SD) achieves optimal performance even with $\mathbb{S} \subset O^{N_t}$ by applying an adaptable search constraint in a sequential manner \cite{fincke1985improved,Agrell02,Damen03,SD}. The SD transforms Eq.~\ref{eqn:ml} into an equivalent tree search 
and applies tree pruning,
visiting fewer nodes and leaves without loss of optimality. 
However, since it is an exact algorithm (\emph{i.e.}, achieves ML performance), the search space for SD is still exponentially large in the worst case~\cite{SDComp1}. Further, because of its sequential nature, its processes cannot be fully parallelized and the complexity (latency) varies per detection, which is not desirable for hardware implementation.

\section{Related Work}
\label{s:relatedwork}

In this section we introduce related work on MIMO detection.






\parahead{Parallel Sub-Optimal Architectures.} These approaches divide the optimal SD tree search into parallel tasks in order to make use of hardware containing many processing elements (PEs) such as a GPU
or FPGA \cite{flexcore-nsdi17,nikitopoulos2018massively}, while search algorithms become approximate. 
For these methods such as the Fixed-Complexity Sphere Decoder (FCSD)~\cite{FCSD,barbero2008extending,jalden2009error,barbero2008low} and K-best SD~\cite{Guo06,mondal2009design,li2008reduced,zheng2017k}, $\mathbb{S}$ is a subset of $\mathcal{O}^{N_t}$, 
so how to select $\mathbb{S}$ for comparing $\mathcal{D}(\mathbf{v})$ is a key factor.  
For instance, 
the FCSD splits the SD tree of $N_t$ levels into two separate search areas, one for \emph{full search} (FS) and the other for \emph{greedy search} (GS). During the FS, the FCSD visits all nodes at the first $N_{fs}$ levels and then switches to GS, where only one child node with minimum partial Euclidean distance is explored for the remaining levels ($N_t-N_{fs}$). This exploration process can run in parallel.\footnote{For the maximum effect of the algorithm, a channel ordering scheme is used to ensure users with poor channel are detected in the FS phase of the FCSD.}
The FCSD results in $\mathbb{S}$ consisting of $|\mathcal{O}|^{N_{fs}}$ candidate solutions. 
Here, $N_{fs}$ is a controllable positive integer parameter that trades off the FCSD's detection performance with its computational complexity. Note that
the complexity of the FCSD (even with small $N_{fs}$) is still larger than linear methods and the FCSD enables only $|\mathcal{O}|^{N_{fs}}$ parallel processes such as 16, 256, 4096 for 16-QAM with $N_{fs}=1,2,3$, respectively (\emph{i.e.,} bounded complexity but not flexible).
\systemname{} features flexible and scalable parallelism. 

\parahead{Heuristics for MIMO Detection.} Heuristic approaches inspired by 
Biology or Combinatorial Optimization methods such as genetic algorithms, 
reactive tabu search, and particle swarm optimization for MIMO detection exist
\cite{abrao2010s, vsvavc2013soft,hedstrom2017achieving}, but significant
performance gains are not observed.  
Some studies have used analog quantum hardware 
platforms \cite{kim2019leveraging,kim2020towards}, but these are specialized platforms that
not yet generally available.
Other studies on Physics-inspired 
SA, Gibbs distribution, and quantum search algorithm show feasibility to some extent~\cite{guangda2009multi,hansen2009near,farhang2006markov,abrao2010s,botsinis2013quantum}, 
but lack comprehensive evaluations for Large and Massive MIMO 
systems and comparisons against other state of the art detectors. 

\section{Design}
\label{s:design}

In this section, we describe the design of \systemname{}: 
Section~\ref{s:PMIS} introduces the key building block of \systemnames{} design,
a SA\hyp{}parallel tempering solver.
Section~\ref{s:system_archtitecture} describes the complete
design of \systemname{}.  Section~\ref{s:2R-paramax} then 
introduces a refinement of \systemname{}, \emph{2R-\systemname{}}, which uses soft
information to enhance performance at the cost of some computational
complexity.  We evaluate both designs in Section~\ref{s:eval}.
\vspace{-0.2cm}
\subsection{ParaMax Ising Solver (PMIS)}
\label{s:PMIS}

The \emph{ParaMax Ising Solver} (PMIS) is the main solver 
module in \systemname{}.  It is based on SA, featuring a parallel tempering
algorithm highly\hyp{}tailored to optimize the Ising model of 
MIMO detection.
PMIS is a completely classical algorithm that does not 
require any specialized hardware for its implementation. 
It performs a local search by updating the spin values of a given 
random initial configuration according to Eq.~\ref{eq:metropolis}. Each 
replica is associated with a different temperature, and temperatures 
may be exchanged according to the update rule in Eq.~\ref{eq:pt}. Since 
the calculation of the energy associated with each replica can be trivially
reduced to matrix\hyp{}vector and vector\hyp{}vector multiplications, 
couplings $g_{ij}$ and local fields $f_j$ of the Ising cost
function $\mathcal{H}$ are stored as a matrix $\mathbf{G}$ and as
a vector $\mathbf{f}$ respectively.
Therefore, the calculation of $\mathcal{H}$, critical for the update
rules in Eq.~\ref{eq:metropolis} and Eq.~\ref{eq:pt}, is reduced to: 
\begin{equation}\label{eq:ising-matrix}
    \mathcal{H}(\mathbf{s}) = \mathbf{s} \cdot \left[\mathbf{G} 
        \cdot \left( \mathbf{s} + 2\mathbf{f} \right) \right]/2,
\end{equation}
with $\mathbf{s}$ the vector representing the spin configuration, 
where the factor $2$ takes into account the symmetry of the matrix 
$\mathbf{G}$. During our implementation, PMIS is optimized to maximize
the performance for operations involving $N_V \lesssim 512$ spin 
variables which cover up to 512, 256, and 128 single\hyp{}antenna 
users with BPSK, QPSK, and 16\hyp{}QAM modulations, respectively. 
We provide further details on our PMIS implementation 
in Section~\ref{s:implementation}.   

\begin{figure}
\includegraphics[width=0.84\linewidth]{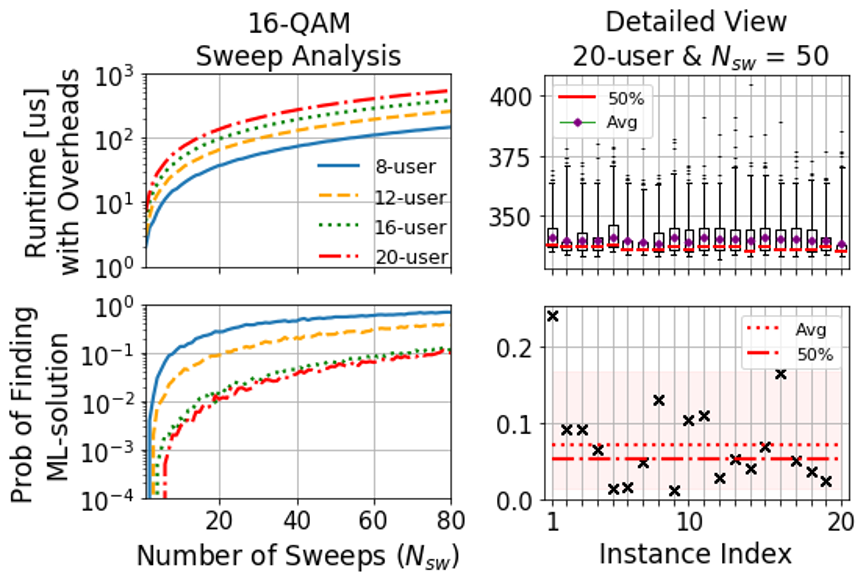}
\caption{\normalfont Metropolis Sweep analysis of the PMIS solving MIMO detection (16-QAM at 20~dB SNR): \emph{\textbf{overview (left)}} varying user numbers and \emph{\textbf{detailed view (right)}}.}
\label{f:sweep_opt}
\end{figure}

\begin{figure}
\includegraphics[width=0.6\linewidth]{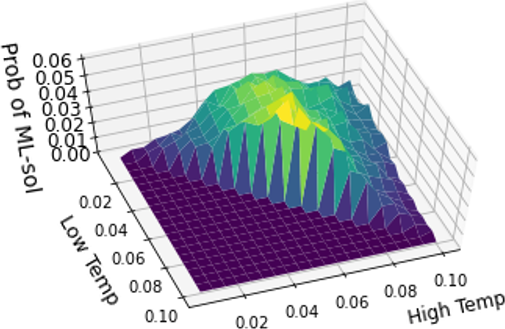}
\caption{\normalfont Temperature range analysis of the PMIS solving MIMO detection for 20-user (16-QAM at 20 dB SNR).}
\label{f:temp}
\end{figure}

\parahead{Computational complexity.} In MIMO detection, 
compute time complexity is a fundamental metric, along with 
BER and network throughput. From Eq.~\ref{eq:ising-matrix}, it 
is clear that complexity is proportional to the square of the number 
of spin variables $N_V$ ($=N_t\,\log_2{|\mathcal{O}|}$). 
Therefore, recalling that every replica is independently updated, 
overall PMIS complexity scales as $N_V^2 \times N_\text{repl} 
\times N_{sw}$, with $N_\text{repl}$ and $N_\text{sw}$ the 
number of replicas and Metropolis sweeps, respectively.

\parahead{Replicas and Metropolis Sweeps.} To reduce the computational 
cost to the bare minimum, we have opted for a ``bang-bang'' parallel 
tempering approach. That is, only two replicas are used ($N_\text{repl}=2$): 
one at very low temperature and one at higher temperature: the replica 
at lower temperature acts as a \emph{greedy} searcher while the 
replica at a higher temperature acts as an \emph{observer}. When 
the greedy searcher is stuck in a local minimum, the two replicas 
can exchange roles (\emph{i.e.}, temperature) to resolve the bottleneck. 
While this has been successfully used in the context of 
quantum annealing \cite{yang2017optimizing}, ours is one of the first
reports of a bang\hyp{}bang parallel tempering schedule on a classical platform.
To choose the number of Metropolis sweeps (\S\ref{s:primer_sa}),  
we empirically examine different numbers of sweeps from one to 80 with 
16-QAM in Figure~\ref{f:sweep_opt} out of 20 instances and 10,000 PMIS 
runs per instance.
Not surprisingly, we observe a trade\hyp{}off between latency (\emph{upper}) and 
sampling quality (\emph{lower}). We choose $N_{sw} = 50$ as an appropriate
point that satisfies the current LTE standard's latency requirements.


\parahead{Choice of temperature range.} 
Unlike $N_\text{repl}$ and $N_\text{sw}$, the temperature range does 
not influence \systemnames{} complexity, so we choose a PMIS temperature 
range where it achieves the highest probability of finding the ML 
solution (\emph{i.e.,} the ground state). For benchmarks we 
select the values $T_\mathrm{min}=0.05$ and $T_\mathrm{max}=0.06$, which 
perform well, as shown in Figure~\ref{f:temp}.


\parabreak{}The foregoing description has described a single PMIS run. 
In \systemname{}, multiple PMIS runs on multiple PEs in parallel, one PMIS run per PE. 
Each PMIS run is independent from the others, accepting
the Ising model $\mathcal{H}$ of the MIMO detection as input,
and outputting a candidate solution. 

\begin{figure}
\centering
\includegraphics[width=0.90\linewidth]{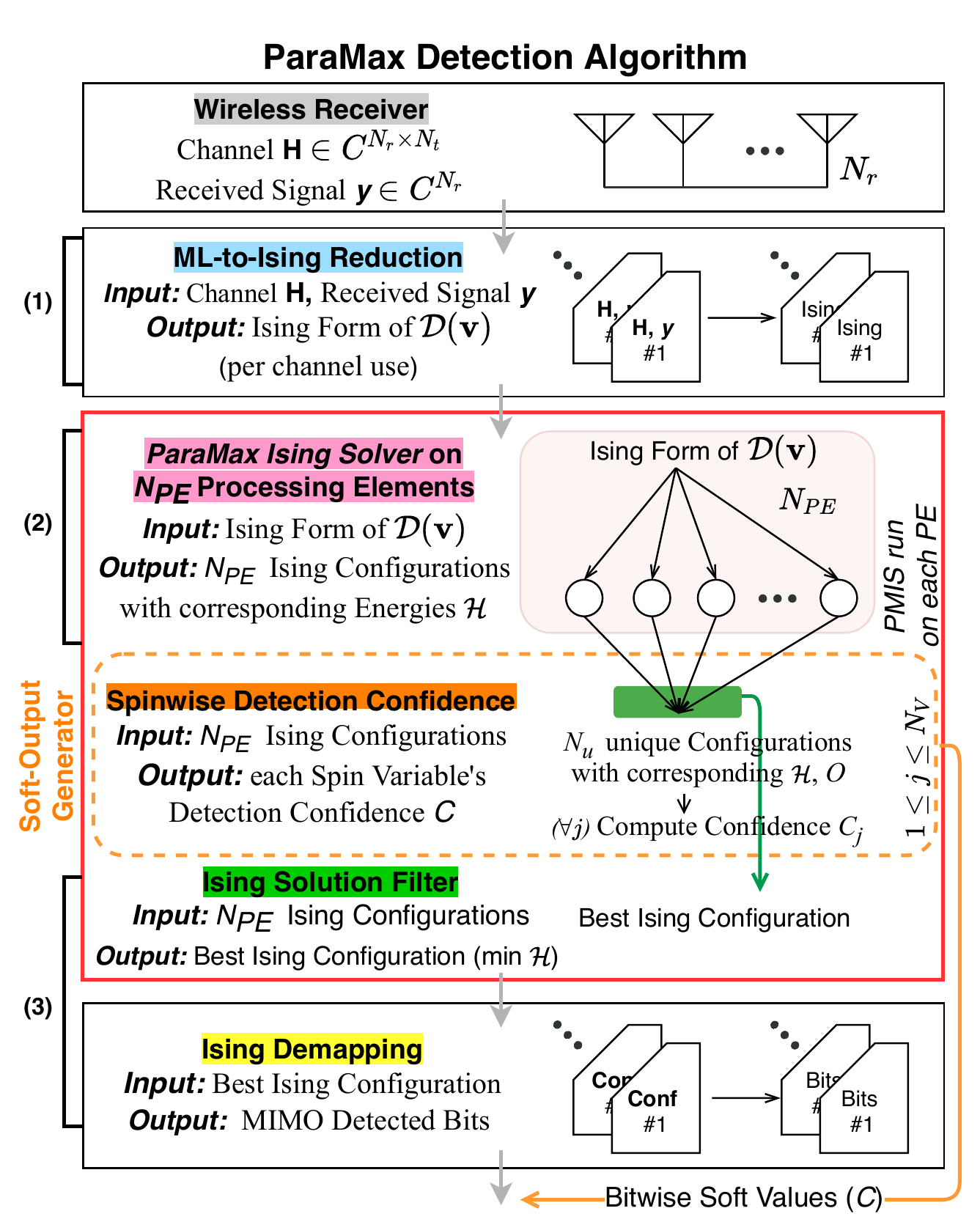}
\caption{\normalfont Overview of \systemnames{} detection algorithm.}
\label{f:paramax}
\end{figure}
\vspace{-0.1cm}

\subsection{\systemname{} Design}
\label{s:system_archtitecture}

In this section we describe the complete \systemname{} design
(Figure~\ref{f:paramax}). We describe the function of each block 
required for MIMO detection in \S\ref{s:paramax_detection} and 
the soft output generator module in \S\ref{s:detection_confidence}.

\subsubsection{\systemname{} Detection Algorithm}
\label{s:paramax_detection}

We assume the base station receives a signal 
perturbed by AWGN and estimates the wireless channel
as stated in Section~\ref{s:primer_mimo_detection}. 

\parahead{1.~ML-to-Ising Reduction.} 
The procedure and the generalized formula of reducing the MIMO 
detection $\mathcal{D}(\mathbf{v})$ 
($= \left\lVert\mathbf{y} - \mathbf{Hv}\right\rVert^2$ 
from Eq.~\ref{eqn:ml}) to the Ising form $\mathcal{H}(\mathbf{s})$ 
using the \emph{spin-to-symbol mapping} were first introduced 
in \cite{kim2019leveraging}; our system assumes the same 
mapping. In the mapping, $N_V$ spins represent all 
possible $N_t$ symbol combinations (\emph{i.e.}, $\log_2{|\mathcal{O}|}$ 
spins for a possible symbol per user),
so its ground state always corresponds to the ML solution.

\parahead{2.~PMIS Parallel Processing.}
Each PMIS run samples an independent solution candidate 
(\emph{i.e.,} an Ising configuration $\mathbf{s}$).  Since detection 
is based on a heuristic,
a single PMIS run's solution may not be optimal, and thus multiple 
runs are required to form a set of candidate solutions, gradually 
increasing the probability of collecting the optimal ML solution.
Since each PMIS optimization run on a single PE completely independent of the
others, \systemname{} flexibly operates on any number of independent 
processing elements ($N_{PE}$), with highly scalable parallelism. 
The number of available processing elements
$N_{PE}$ is equal to the number of PMIS outputs that the \systemname{} 
system can generate with full parallelism.

\parahead{3.~Ising Solution Filter and Demapping.}
After all $N_{PE}$ PMIS parallel runs, the corresponding $N_{PE}$ outputs 
are collected in a table of Ising spin configurations.
Before further processing, the list is sorted in order of solution quality, 
based on the Ising energy $\mathcal{H}(\mathbf{s})$ of each output. 
The \emph{Ising solution filter} returns only the configuration 
$\mathbf{\hat{s}}$ with the best (minimum) $\mathcal{H}(\mathbf{\hat{s}})$, which 
is equivalent to the wireless symbol $\hat{\mathbf{v}}$, \emph{i.e.,} $\mathbf{v}$ with the minimum $\mathcal{D}(\mathbf{v})$ (among candidate solutions), after proper 
demapping \mbox{(spins $\rightarrow$ symbols)}. Finally, $\mathbf{\hat{v}}$ is converted into $N_V$ MIMO detected bits.

\subsubsection{Spinwise Soft Information Output}
\label{s:detection_confidence}

For most heuristics\hyp{}based solvers, only the lowest\hyp{}energy Ising configuration
is returned (regardless of how many times it occurs among $N_{PE}$ PMIS outputs) 
and any outputs other than it are discarded. In \systemname{}, however,
we utilize all $N_{PE}$ PMIS outputs to generate soft information
(\emph{i.e.,} detection confidences, for each spin in a given configuration). 
In general, soft\hyp{}output MIMO detectors' soft values are 
utilized for iterative MIMO detection or channel coding \cite{roger2012efficient,larsson2002maximum,larsson2008fixed,barbero2008low}. 
In this work, we design the former (2R-\systemname{}) in Section~\ref{s:2R-paramax}. 

\begin{figure}
\centering
\includegraphics[trim={0 0 1.5cm 0},width=0.95\linewidth]{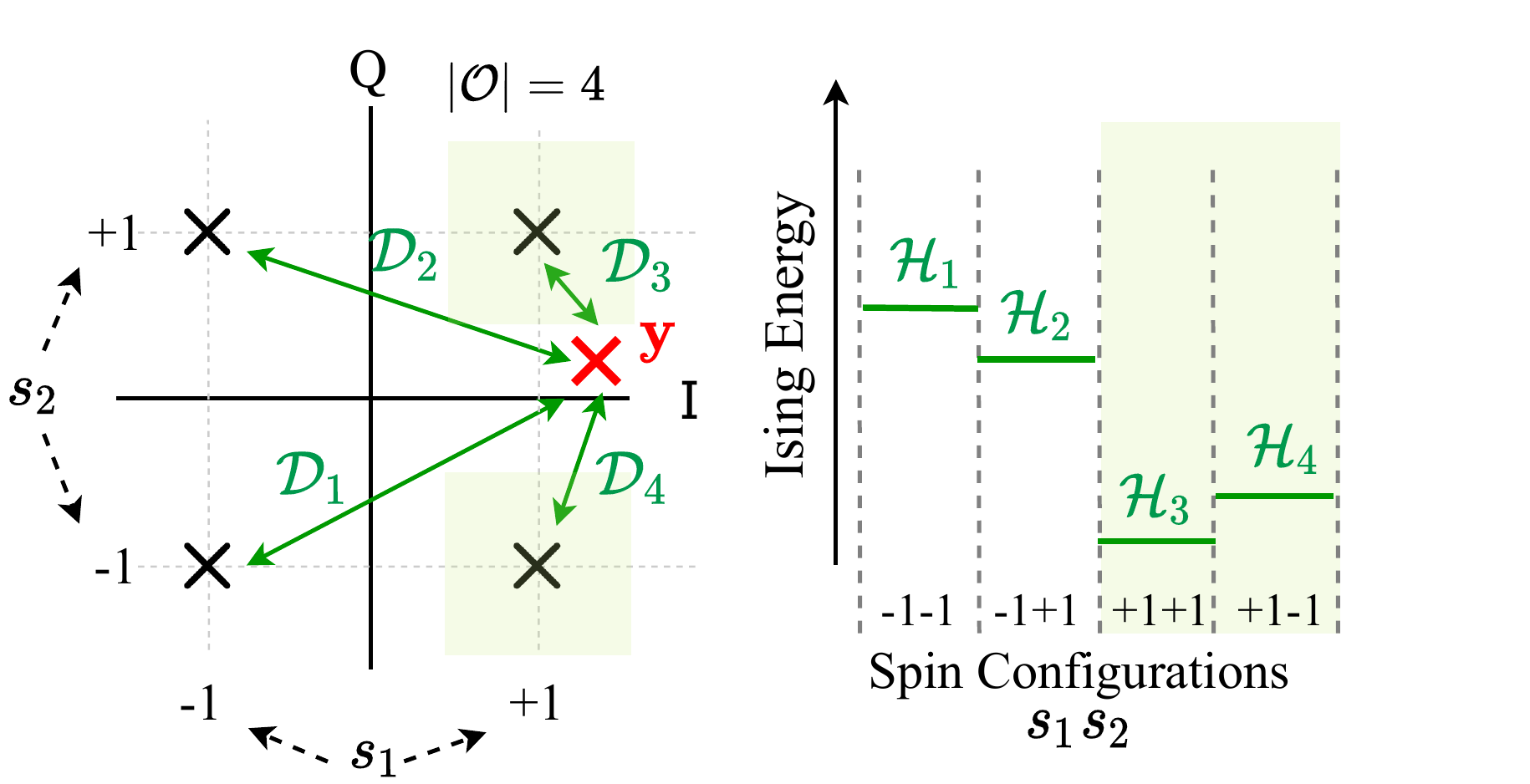}
\caption{\normalfont Equivalent representations of 1$\times$1 QPSK detection:
Euclidean distances $\mathcal{D}(\mathbf{v})$ in the I-Q plane \emph{(left)},
and Ising energies $\mathcal{H}(\mathbf{s})$ \emph{(right)}. 
Shadings highlight likely solutions.}
\label{f:ml_to_isng}
\end{figure}

\systemname{} collects candidate solutions from $N_{PE}$ independent
PMIS runs. Among these, multiple occurrences of a certain spin 
configuration (with agreeing spin variables) are very likely to be observed, 
which could be used to identify spins easy (or hard) to detect (\emph{i.e.} 
variables that are very likely to be assigned a certain value in 
the unknown optimal ML solution). Figure~\ref{f:ml_to_isng} shows 
an illustrative example of detecting a received $1\times 1$ QPSK signal $\mathbf{y}$ in
two equivalent representations, one in the I-Q plane with 
Euclidean distance $\mathcal{D}$ \emph{(left)}, and the other with 
Ising energies $\mathcal{H}$ \emph{(right)}. 
In this example, the first spin variable $s_1$ (corresponding
to the symbol's real part) is likely to be detected as $+1$ for
most PMIS runs, since the difference in Ising energy from all 
configurations that have $s_1=+1$ (resulting in $\mathcal{H}_3$ or $\mathcal{H}_4$ 
in Figure~\ref{f:ml_to_isng}, \emph{right})
and $s_1=-1$ (resulting in $\mathcal{H}_1$ or $\mathcal{H}_2$ in 
Figure~\ref{f:ml_to_isng}, \emph{right}) is significant.
The spin $s_1$ is easy to detect compared to spin $s_2$ (corresponding
to the symbol's imaginary part).  Multiple occurrences of PMIS runs
agreeing on $s_1=+1$ indicate this, while PMIS runs will 
disagree on the value of $s_2$, because the two 
most frequent spin configurations' energies ($\mathcal{H}_3$ and $\mathcal{H}_4$)
themselves disagree on the value of $s_2$.

\begin{figure}
\centering
\includegraphics[width=0.85\linewidth]{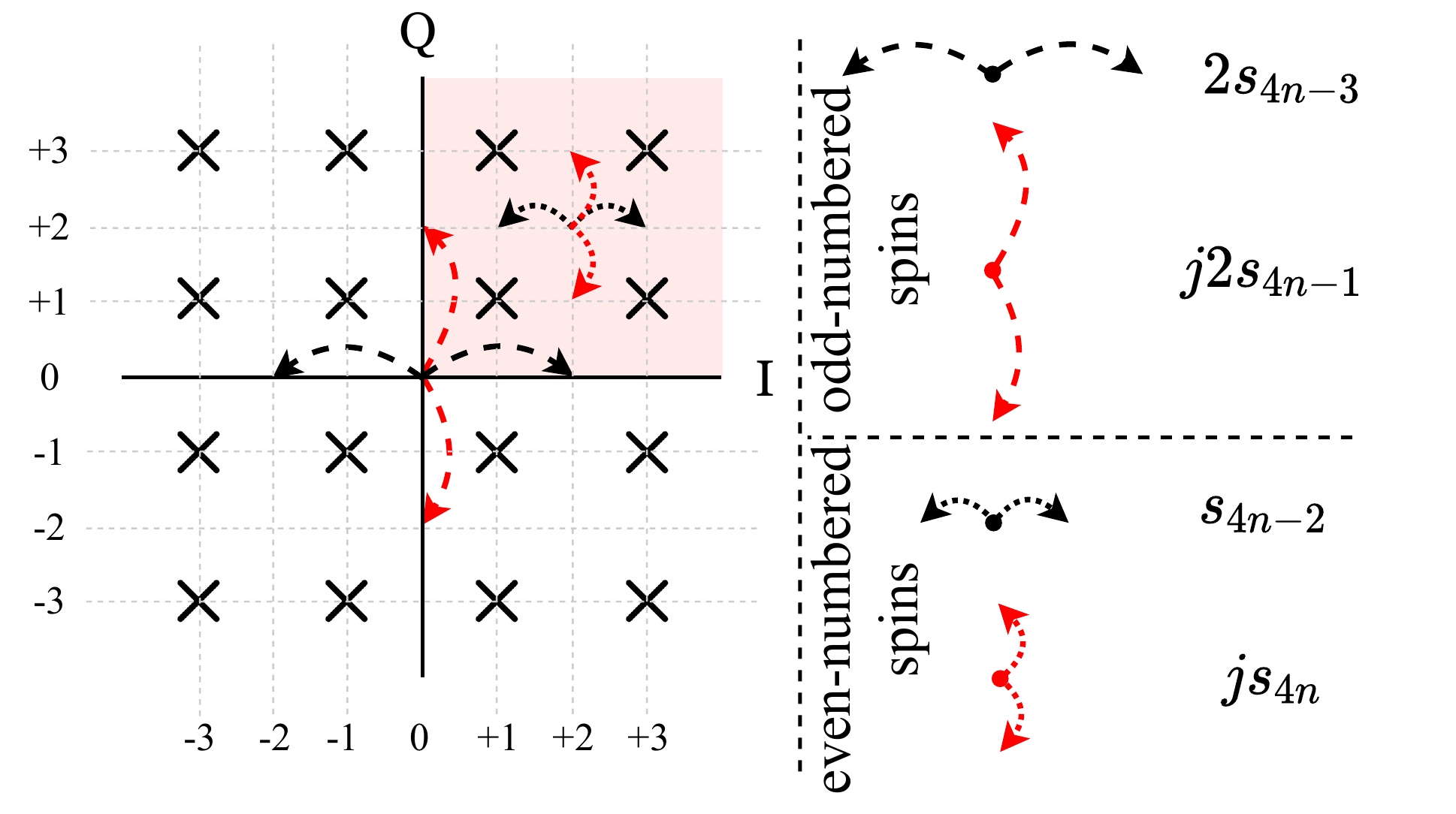}
\caption{\normalfont Mapping between four Ising spins (variables) to n-th user's 16-QAM symbols ($\times$ symbols) on the constellation. Shading denotes an inner quadrant. In this mapping, odd numbered spins are easier to detect than even numbered spins.}
\label{f:16qam_const}
\end{figure}

This phenomenon becomes even  clearer for high\hyp{}order modulations, 
since in 16-QAM or higher modulations, the value of the spin coefficients 
in the Ising spin\hyp{}to\hyp{}symbol mapping varies across spins.
For example, Figure~\ref{f:16qam_const} shows the spin\hyp{}to\hyp{}symbol
mapping of 16-QAM for the $n^{\mathrm{th}}$ user, where the
user's possible symbol maps one\hyp{}to\hyp{}one with spins
$s_{4n-3}, \ldots, s_{4n}$, that is
$\mathbf{v_n} = (2s_{4n-3}+s_{4n-2}) + j(2q_{4n-1}+s_{4n})$. 
Here, spins $s_{4n-3}$ and $s_{4n-1}$, the odd\hyp{}numbered spins,
determine the symbol's quadrant, while spins $s_{4n-2}$ and $s_{4n}$,
the even\hyp{}numbered spins, determine the symbol in the given 
quadrant. Here, the odd\hyp{}numbered spins for 16-QAM are in 
general easier to detect than the even\hyp{}numbered spins 
because of higher robustness to AWGN (detection reliability).
Table~\ref{t:error_case} presents empirical spinwise error rates
of \systemname{} for $8\times8$ 16-QAM detection.
These differences in robustness indicate that using \systemnames{} 
soft information would be particularly helpful for further 
processing, similarly for unequal error protection (UEP) 
\cite{masnick1967linear,wei1993coded,horn1999robust,boyarinov1981linear,237878}). 


\begin{table}[htbp]
\centering
\caption{\normalfont \systemnames{} spinwise error rate
(conditioned on 103,850 incorrect outputs)
for eight\hyp{}user, 16\hyp{}QAM MIMO detection.}
\begin{tabular}{ c c c c}
\multicolumn{4}{c}{\textbf{Mean Spinwise Error Rate}}\\
\toprule
\multicolumn{2}{c}{\textbf{Oven\hyp{}numbered spins:}} & \multicolumn{2}{c}{\textbf{Even\hyp{}numbered spins:}} \\
 \textbf{$4n-3^{\mathrm{rd}}$} & \textbf{$4n-1^{\mathrm{st}}$} & \textbf{$4n-2^{\mathrm{nd}}$} &  \textbf{$4n^{\mathrm{th}}$} \\
\hline
 0.167 &  0.152 &  0.329 &  0.352  \\
 \multicolumn{2}{l}{$\approx$ Either/both: 0.32} & \multicolumn{2}{l}{$\approx$ Either/both: 0.68} \\  
\bottomrule
\end{tabular}
\label{t:error_case}
\end{table}

\parahead{Soft information computation.} Based on the equivalence of 
of the I-Q and spin configuration representations, the occurrence count 
of a given spin value for a certain spin $s_j$ across all 
PMIS outputs samples the distance of the symbol corresponding to 
that spin's value, and hence estimates that spin's likelihood 
of correctness.
More specifically, after collecting $N_{PE}$ PMIS outputs sorted
by Ising energies $\mathcal{H}_i$ ($1 \leq i \leq N_u$), the
system has $N_u$ ($N_u \leq N_{PE}$) unique outputs with 
corresponding \emph{occurrence counts} $O_i$ ($1 \leq i \leq N_u$). 
The detection confidence $C_{j}$ of spin $s_j$ ($1 \leq j \leq N_V$) 
is defined as:
\begin{align}
C_{j} = \left(\sum^{N_u}_{i=1} O^{s^i_j=s^{1}_j}_{i} \cdot \abs{\frac{\mathcal{H}_i}{\mathcal{H}_1}}\right)\Big/\left(\sum^{N_u}_{i=1} O_{i}\cdot \abs{\frac{\mathcal{H}_i}{\mathcal{H}_1}}\right), 
\label{eqn:detection_confidence}
\end{align}
where $O^{s^i_j=s^{1}_j}_i$ is a count of 
occurrences of the $i^{\mathrm{th}}$
ranked configuration (defined in \S\ref{s:primer_sa} on p.~\pageref{s:primer_sa}),
only when the $i^{\mathrm{th}}$
configuration's $j^{\mathrm{th}}$ spin is equal to the first\hyp{}ranked
configuration's $j^{\mathrm{th}}$ spin (\emph{i.e.},  $O^{s^i_j=s^{1}_j}_i$ 
is either $O_i$ or zero). 
The spinwise detection confidences $C_j$ ($0 < C \leq 1.0$) are the 
soft values \systemname{} outputs in this step. 
Note that the reliability of each $C_j$ increases as the best observed
Ising energy among the collected $N_{PE}$ outputs ($\mathcal{H}_1$) 
becomes closer to the unknown ground state (of energy $\mathcal{H}$), 
which implies as $N_{PE}$ increases the quality of soft values 
improves.\footnote{While \systemname{} is an inherent soft\hyp{}output MIMO detector, 
requiring simple computations, conventional soft-output MIMO detectors 
require additional computations of exponential complexity, of log\hyp{}likelihood 
ratio (LLR) for all coded bits to generate soft values at channel decoder~\cite{roger2012efficient,studer2006soft,wang2004approaching}.} Similar algorithm is introduced using quantum 
annealing~\cite{karimi2017boosting}, where only partial outputs are used. 
\vspace{-0.2cm}

\subsection{2R-\systemname{}: Iterative Soft Detection}
\label{s:2R-paramax}

\begin{figure}
\centering
\includegraphics[width=0.95\linewidth]{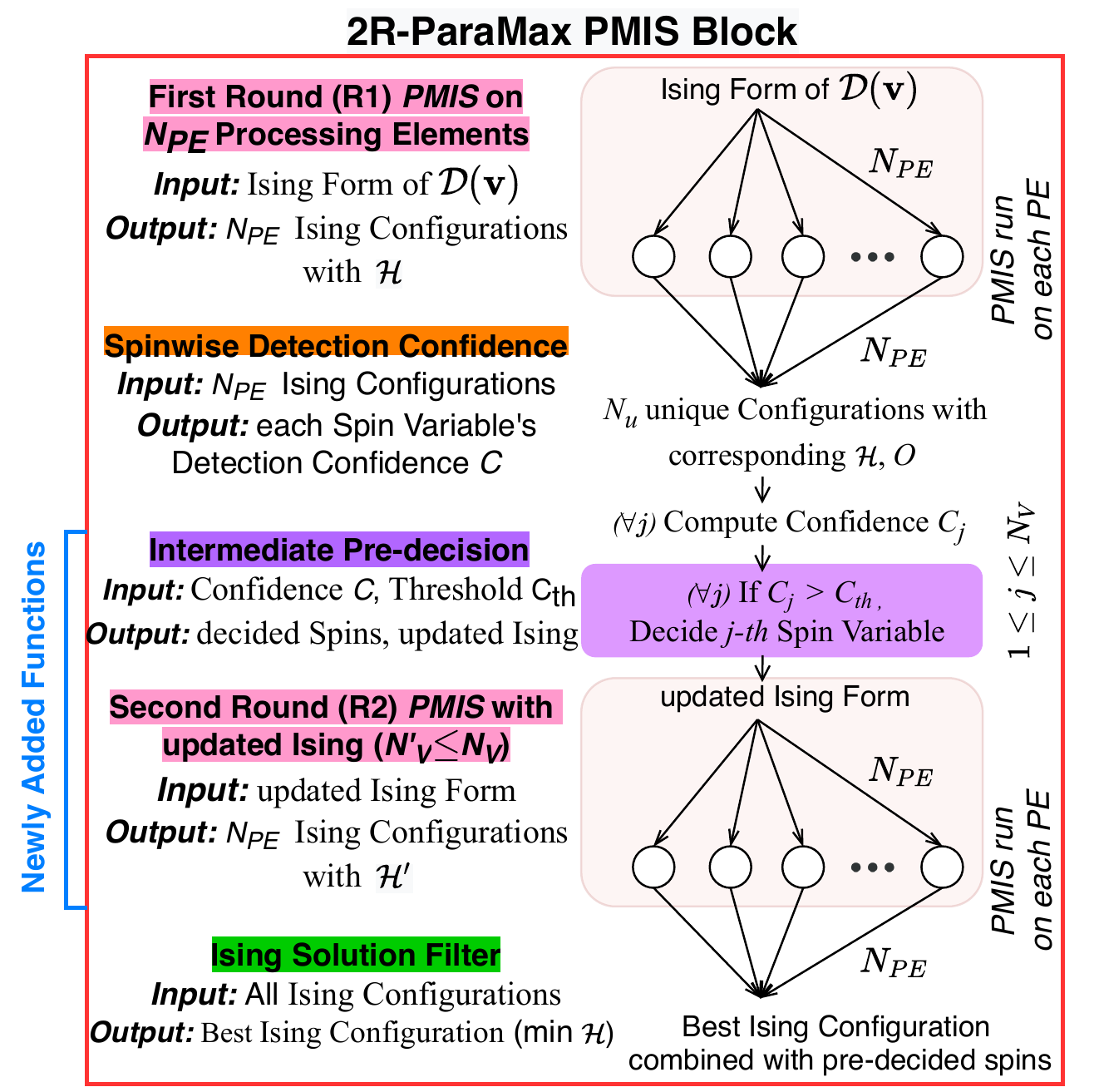}
\caption{\normalfont Structure of 2R-\systemname{} PMIS Block (\emph{cf.} Figure~\ref{f:paramax}'s third block in red). The other blocks in 2R-\systemname{} are exactly the same as ones in \systemname{}.}
\label{f:2R-paramax}
\end{figure}

We now introduce a method of using the soft information described in
the prior section to enhance the operation of \systemname{}.  We call 
this protocol 2R-\systemname{}.  The main idea is to iterate the 
PMIS block twice, once for generating soft confidence information, and again
to obtain a final detection result based on the confidences from the first
iteration.  Intermediate processing between the first and second iterations
functions pre-decision of spins with high detection confidence.
An error correction post-processing is applied at the end of the second round,
both of which have linear complexity. The end result is a more accurate 
MIMO detection result, at the expense of a modestly increased latency,
and so this might be employed for challenging wireless channels and\fshyp{}or
large numbers of users. With reference to Figure~\ref{f:paramax}, the structure of 2R-\systemname{} PMIS block is 
shown in Figure~\ref{f:2R-paramax}. This block is replacing the third 
block marked in red of Figure~\ref{f:paramax}. 
The other blocks are exactly the same as described in \systemname{}. We also note that the soft information generated by the second round 
can also be used for the channel decoding or further iterations of the 
algorithm. 

\parahead{Intermediate Pre\hyp{}decision.} The intermediate pre\hyp{}decision 
module identifies those spins with a high detection confidence 
(over a threshold $C_{\mathrm{th}}$) from the first round 
of PMIS outputs in order to reduce the number of spin variables involved
in second\hyp{}round of PMIS runs, simplifying second\hyp{}round detection.
That is, if the $j^{\mathrm{th}}$ spin's detection confidence 
$C_j \geq C_{\mathrm{th}}$ ($1 \leq j \leq N_V$), then we pre\hyp{}decide 
the $j^{\mathrm{th}}$ spin variable to be the value of the corresponding 
spin of the best solution in the first round (\emph{i.e.}, $s^1_j$). 

After thus obtaining a pre\hyp{}decided set of spin indices, the next step is to update the Ising form accordingly.
For each spin index $k$ in $\mathcal{F}$ and for each Ising problem
index $i$, we set 
$f'_i = f_i + g_{ik}\cdot s^1_k,\,\,\text{if } i < k\, \,\, 
    \text{and}\,\, f'_i = f_i + g_{ki}\cdot s^1_k, \,\, 
    \text{if } i > k$,
and then remove $f_k$, $g_{ik}$ and $g_{ki}$. 
The result is a reduced Ising problem that contains $N_V^\prime=N_V-|\mathcal{F}|$ spins only.
Table~\ref{t:pre_decision} summarizes the average ratio of decided spins ($|\mathcal{F}|/N_V$) and the success ratio ($|\mathcal{F_{\text{ML}}}|/|\mathcal{F}|$) of the pre-decision process.
Here, $\mathcal{F}_{\text{ML}}$ denotes an index group of spins where spins decided by the pre-decision process are exactly the same as the corresponding spins in the ML solution. In 2R-\systemname{}, we apply $C_{th}\geq 0.97$, which ensures   $|\mathcal{F_{\text{ML}}}|/|\mathcal{F}|=1.0$ (for lower $N_{PE}$, higher $C_{th}$ applied). With the updated Ising form $\mathcal{H'}$ (with $f'$ and $g'$ for $N_V'$ spins), we execute a second round of PMIS to generate $N_{PE}$ outputs and then filter the best output consisting of $N_V'$ spins with minimum Ising energy in terms of $\mathcal{H'}$. When the filtered configuration is combined with the pre-decided spins appropriately, the full configuration consisting of $N_V$ spins can be restored and demapped into symbols. This full configuration is further compared against the best PMIS outputs of the first round based on the original Ising form $\mathcal{H}$. The final best configuration is returned as 2R-\systemname{}'s detection solution, which guarantees that 2R-\systemnames{} bare minimum performance is \systemnames{} performance.
\vspace{-0.1cm}

\begin{table}[htbp]
\centering
\caption{\normalfont 2R-\systemnames{} intermediate pre-decision process tested for  5,000 different instances of 20 $\times$ 20 16-QAM detection at SNR 20~dB on $N_{PE}=200$.}
\begin{tabularx}{\linewidth}{X*{5}{@{\hskip 12pt}r}}
\toprule
$\mathbf{C_{th}}$ & \textbf{0.91} &\textbf{0.93} & \textbf{0.95} & \textbf{0.97} &\textbf{0.99} \\
\midrule
$\mathbf{|\mathcal{F}|/N_V}$ & 0.35 &0.32 & 0.28 & 0.21 & 0.01 \\
$|\mathcal{F_{\text{ML}}}|/|\mathcal{F}|$ & 0.97 & 0.99 & 1.0 & 1.0 & 1.0 \\
\bottomrule
\end{tabularx}
\label{t:pre_decision}
\end{table}
\vspace{-0.4cm}

\section{Implementation}
\label{s:implementation}

We now describe our \systemname{} implementation. 

\parahead{Computing Environments.} CPU-based experiments are executed on
an Intel i9-9820X at 3.30GHz  with 20 cores, 2,189 threads, and 126~GB
RAM. GPU-based experiments are tested based on the CUDA (Compute
Unified Device Architecture~\cite{sanders2010cuda}) 10.2 with GeForce 
RTX 2080 Ti of 4,352 CUDA cores and 68 streaming multiprocessors. 

\parahead{Wireless MIMO Channels.} Both simulation\hyp{}based and 
trace\hyp{}driven real world wireless channels are used for 
our experiments. In the case of the simulation\hyp{}based channel, 
independent and identically distributed (i.i.d) Gaussian channels with
AWGN are synthesized for various SNR settings. For trace\hyp{}driven channels,
we use non\hyp{}line of sight wideband MIMO channel traces at 2.4~GHz, 
between 96 base station antennas ($N_r$) and eight static users ($N_t$), 
the largest MU-MIMO dataset provided in Argos~\cite{Argos}. 
Among $N_r=96$, we single out 8 to 32 (in steps of four) antennas
to test the most challenging MIMO regimes (\emph{e.g.} $N_t \geq N_r/4$).
Since trace based channels include measured noise and limited user numbers,
we use synthesized channels, unless otherwise stated, in order 
to precisely control SNRs and evaluate various MIMO regimes 
such as $N_t > 8$. Based on both channel settings, we generate 
large-scale Ising models $\mathcal{H}$ of MIMO detection 
(100,000\hyp{}1,000,000 random instances per scenario) in 
order to measure detection BER up to $(N_V\cdot\text{total Insts})^{-1}$,
approximately $10^{-7}$.

\parahead{PMIS CPU-Implementation.}
While the front-end of our PMIS implementation is in Python, the core is completely written in $C^{++}11$ standard~\cite{josuttis2012c++}. We assign only a single core and a single thread (a single PE) to the calculation of each PMIS run by manually modifying the OpenMP~\cite{chandra2001parallel,dagum1998openmp} and C++ parallelization settings.
Furthemore, to maximize the performance of PMIS (to satisfy limited processing time in wireless standards), the following \emph{innovations} have been implemented: \textbf{(1) Use of static memory:} Static allocation, unlike dynamic allocation, happens at global scope and it is pre-populated when the library is loaded. Moreover, since the size of arrays is known in advance, compilers can further optimize math operations on static arrays. \textbf{(2) Parameter pack expansion:} Loops in the matrix-matrix and vector-matrix multiplications are the most expensive part in PMIS implementation. To further reduce the computational cost, most of the critical loops are statically unrolled using features like the parameter pack expansion, introduced in the $C^{++}11$ standard. \textbf{(3) Intel SIMD instructions:} Most of the modern CPU architecture have intrinsic operations to allow multiple operations on contiguous arrays of floats. In PMIS, we have used Intel SIMD instructions to vectorialize operations like matrix-vector and vector-vector multiplications~\cite{plank2013screaming}. 

\parahead{PMIS GPU-Implementation.}
The core design of CPU-\systemname{} is based on a highly
optimized C++ implementation of SA. Therefore, the natural extension
of \systemname{} to GPU consists into the implementation of the core SA engine to GPU.
However, unlike CPUs, GPUs achieve the best performance for large arrays where multiple
synchronous operations are applied at the same time. Indeed, while GPUs have more cores
than CPUs, each single GPU core is typically much slower. Therefore, to maximize
the performance of SA implemented on GPUs, we have designed a GPU kernel based on the JAX/XLA language that updates
multiple PMIS runs at the same time. More precisely, the spin configuration $\mathbf{s}$ for a single
PMIS run (in Eq.~\ref{eq:ising-matrix}) is now extended to a matrix $\mathbf{S}\, (=\mathbf{s}^k)$, with $k$ corresponding to the PMIS index. Since PMIS runs are completely independent 
from each other, $\mathbf{s}^k$ ($\forall k$)
can be updated independently and synchronously.

\section{Evaluation}
\label{s:eval}

In this section, we evaluate \systemname{} in various aspects. Section~\ref{s:eval_latency} evaluates \systemnames{} detection latency against other CPU- and GPU-based detectors. Section~\ref{s:eval_sampling} illustrates sampling performance of \systemname{} comparing against simulated annealing and required the number of processing elements for \systemname{} to achieve near-ML performance in both Large and Massive MIMO.  Section~\ref{s:eval_ber_throughput} and ~\ref{s:eval_throughput} show the \systemnames{} bit error rate and system throughput performance respectively, compared against other state-of-the-art detectors in both Large and Massive MIMO.

\begin{figure}
\centering
 \begin{subfigure}[b]{\linewidth}
    \centering
    \includegraphics[width=0.87\linewidth]{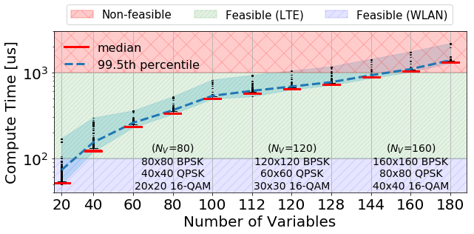}
    
    \caption{\normalfont Compute latency of fully-parallel full-blown C++ kernel CPU-\systemname{}.}
    \label{f:latency}
\end{subfigure}
 \begin{subfigure}[b]{\linewidth}
    \centering
    \includegraphics[width=\linewidth]{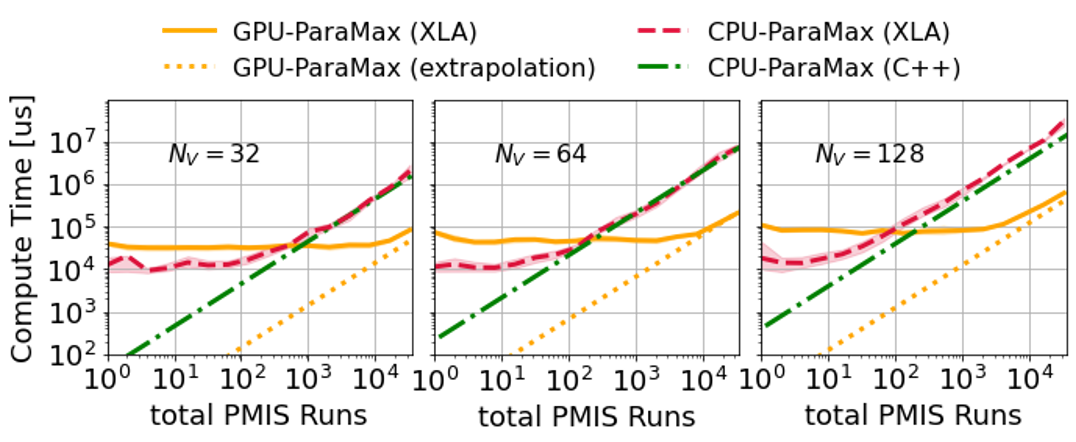}
    \caption{\normalfont Single Device: 9.3~TFLOP GPU-\systemname{} vs 0.028~TFLOP CPU-\systemname{}.}
    \label{f:cpu_vs_gpu}    
\end{subfigure}
\caption{\normalfont \textbf{\systemname{} Detection Latency.} 
}
\label{f:compute_vs_nv}
\end{figure}


\subsection{Detection Latency}
\label{s:eval_latency}

\parahead{Fully-Parallel Full-Blown ParaMax.}
Figure~\ref{f:latency} shows the detection time of \systemname{} as a function of $N_V$ ($= N_t\, \log_2{|\mathcal{O}|})$ per channel use, where the background color coding indicates approximate feasibility for the wireless standards (WLAN and LTE). As $N_V$ increases (\emph{i.e.}, $N_t$ and/or modulation increases), computing time tends to scale as $N_V^2$ (cf. $N_r$ does not affect $N_V$ and thus compute time). The available largest MIMO sizes that reach the borderline-limit of acceptable detecting time 
in the LTE standards are $160\times N_r$, $80\times N_r$, $40\times N_r$ for BPSK, QPSK, 16-QAM modulation, respectively. 
While slight variations of runtime are observed that can cause overall latency increase and hardware synchronization issues, this can be resolved by an integrated system hardware since 
the origin of the variations is
caused 
related to our system
by the kernel allocation of jobs and concurrent
(unrelated) system processes. 
In general, 2R-\systemname{} requires approximately $1.4-1.6${\small $\times$} \systemname{} latency.



\parahead{CPU- vs GPU-ParaMax Comparison.}
Figure~\ref{f:cpu_vs_gpu} 
shows the comparison of compute time between the XLA kernel (compiled for both CPU and GPU) and the original C++ kernel. The runtime for the C++ kernel has been obtained by computing the average runtime for a single PMIS run and then projected for multiple PMIS runs. XLA (Accelerated Linear Algebra) is a domain-specific compiler for linear algebra that can be compiled and optimized separately for either CPU or GPU.
%
%
%
As one can see, for a sufficiently large number of parallel PMIS runs, while the kernel runtime compiled for CPUs have a consistent runtime with the full-blown C++ implementation of ParaMax, the kernel compiled for GPUs shows a speed-up, where total PMIS runs can be defined as $N_{PE}$ multiplied by the number of subcarriers ($N_{SC}$). 
While GPU-\systemname{} can achieve the speed-up for over hundreds of PMIS runs, it cannot satisfy time requirements for standards. Recall that current GPUs are not designed to make full use of resources for small-size systems (\emph{i.e.,} few PMIS runs). Thus we also extrapolate its performance to estimate what we could achieve in GPU without these limitations. Note that unlike on CPUs where a single core can be used to carry out any calculation, GPU cores are designed to work in concert to manipulate large block of data in parallel, and users cannot assign specific resources to a certain computation~\cite{flexcore-nsdi17}. Therefore, we define a single PE for GPU-\systemname{} as the extrapolation of a single PMIS run from large number of PMIS runs. In 5G New Radio, this extrapolation becomes more reasonable (while still being approximate), since 5G systems will support over three thousands of subcarriers, and slightly more time\footnote{Available compute time is for all BS-processing including channel decoding.} (4~ms) than LTE (3~ms) will be allowed for enhanced mobile broadband (eMBB)~\cite{ding2020agora,BigStation,3gpp-ts-36-211} (cf. 1~ms for 5G Ultra-Reliable Low-Latency Communication (URLLC)). 
\vspace{-0.4cm}

\begin{table}[htbp]
\centering
\begin{tabularx}{\linewidth}{{m{0.00000000000000000000001mm}@{\hskip 0.1in}c@{\hskip 0.09in}c @{\hskip 0.07in} c@{\hskip 0.07in} c c c}X*{6}{c}}
    \multicolumn{2}{l}{\textit{\scriptsize 20-user MIMO (16-QAM)}} & \multirow{2}{*}{\textbf{ZF-SIC}} &\multirow{2}{*}{\textbf{ParaMax}} & 
    \multicolumn{3}{c}{\textbf{FCSD}}\\
    [0ex]\cline{5-7} 
  & & {} & &  
 $N_{fs}$=2  &   $N_{fs}$=3&   $N_{fs}$=4 \\[0ex]
\cline{1-7}
\multicolumn{2}{c}{\textbf{\scriptsize Parallelism \#}} 
 &$\times$ & \multirow{2}{*}{\textit{Flexible}} & $16^2$ & $16^3$ & $16^4$\\ 
\multicolumn{2}{c}{\textbf{\scriptsize Required $N_{PE}$}}
 & 
 1 & 
  &
 256 &
 4,096 & 
 65,536\\
\cline{1-7}
\multirow{4}{*}{\rotatebox[origin=c]{90}{\textbf{\scriptsize Latency [us]}}}  & {CPU} &  25  &   357 &405 &  5,821 &  93,714 
\\ 
{} & {GPU} & 83,861 & extr. 31  &   319 &   378 &  1,841\\
\cline{2-7}
{} &  {\scriptsize } & CPU &  CPU  & GPU & GPU & GPU\\
{} &  {\scriptsize \emph{Min} time} & 
\cellcolor{LightGreen} 
\textbf{25} & 
\cellcolor{LightGreen}  
\textbf{357}  & 
\cellcolor{LightGreen}  
\textbf{319}  & 
\cellcolor{LightGreen} 
\textbf{378} & 
\cellcolor{LightRed} 
1,841
\\
\cline{1-7}
\end{tabularx}
\caption{\normalfont
Available number of parallel processes, required $N_{PE}$ for fully parallel processing, and average detection runtime of various MIMO detectors both on CPU and GPU. \systemnames{} compute time is for a single PMIS run on a single PE (\emph{i.e.,} fully-parallel \systemname{}) and GPU-\systemname{} reports extrapolated compute time.}
\label{t:classic_solver}
\end{table}

\vspace{-0.45cm}

\parahead{Comparison against Conventional Detectors.}
We compare \systemname{} latency against various detectors implemented on the MIMOPACK library~\cite{ramiro2015mimopack}, which is one of the fastest open-source MIMO detector implementations based on the (CUDA) C programming. The results for 20-user 16-QAM are summarized in Table~\ref{t:classic_solver}. In the case of the zero-forcing successive interference cancellation (ZF-SIC or V-BLAST with ordering scheme)~\cite{wolniansky1998v}, 
while its complexity is slightly higher than linear detectors such as ZF and MMSE, compute time is still few tens of microseconds. 
However, their computations (both ZF and ZF-SIC) are not appropriate for parallel processing, causing extra overheads such as job scheduling and data transition among computing resources.  
In the case of the FCSD, we consider three different $N_{fs}$ that trade-offs the FCSD's detection performance with its complexity. 
As long as the available number of PEs is large enough to allow full parallelism ($|\mathcal{O}|^{N_{fs}}\leq$ total PEs), the compute time remains in a few hundreds of microseconds, satisfying LTE requirements. 
\vspace{-0.1cm}



\subsection{Heuristic Detection Sampling}
\label{s:eval_sampling}

For \systemname{}, we can report the expected number of sampling repetitions to reach ML-performance, which can be computed using the \emph{probability of obtaining the ML-solution in one sample} of a given MIMO detection scenario (\emph{i.e.,} MIMO size, modulation, and SNR), averaged across the problem distribution ($P_\text{ML}$)~\cite{mandra2016strengths}. Since $P_\text{ML}$ cannot be determined \emph{a priori} by theoretical means, we obtain it through empirical evaluation of statistically significant 1,000,000 PMIS runs across 100 detection instances per scenario. In order to compute this average probability, we use the ML-solutions found by expensive runs of the Sphere Decoder. Since each run is independent, the  probability for \systemnames{} to find the optimal ML-solution: \begin{equation}\label{eq:para-ml}
    \mathcal{P}(\text{ParaMax}_\text{ML}) = 1-(1-P_\text{ML})^{N_{PE}}.
\end{equation}
Inverting Eq.~\ref{eq:para-ml}, we can obtain the required number of PMIS repetitions (samples) to achieve the ML-detection with a target probability $\mathcal{P}_\mathbf{T}({\text{ParaMax}_\text{ML}})$ as:
\begin{equation}\label{eq:n-rep}
    \text{required } N_{PE} = \frac{\log(1-\mathcal{P}_\mathbf{T}(\text{ParaMax}_\text{ML}))}{\log(1 - P_\text{ML})}. 
\end{equation}

\begin{figure}
    \centering
    \begin{subfigure}[b]{\linewidth}
    \centering
    \includegraphics[width=\linewidth]{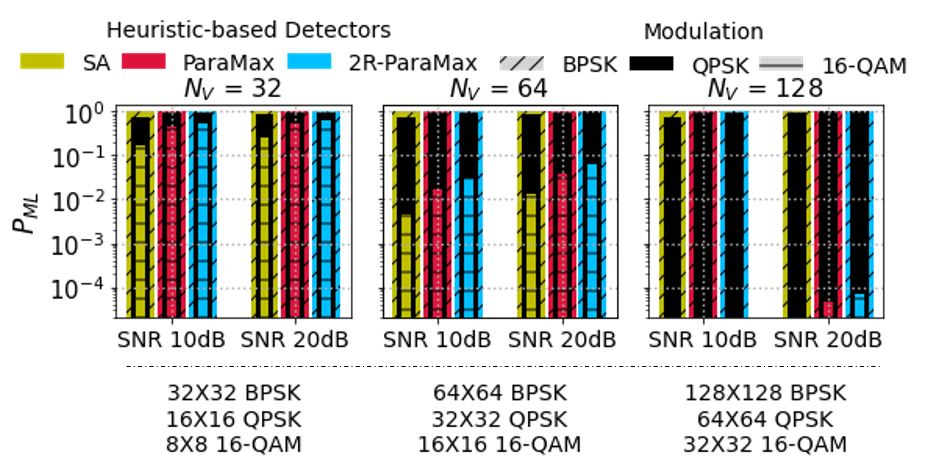}
    
    \caption{\normalfont Average probability of finding the ML-solution per run ($P_\text{ML}$).}
    \label{f:prob_ml_comparison}
    \end{subfigure}
    \qquad
    \begin{subfigure}[b]{\linewidth}
    \centering
    \includegraphics[width=0.95\linewidth]{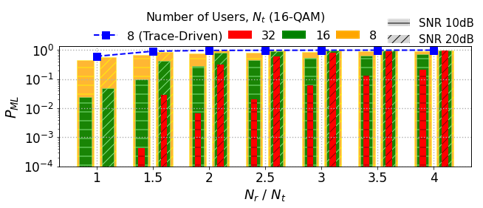}
    \caption{\normalfont Impact of $N_r/N_t$ ratios (from Large  to Massive MIMO).}
    \label{f:prob_ml_varying_ratio}
    \end{subfigure}
    \qquad
    \begin{subfigure}[b]{\linewidth}
    \includegraphics[width=0.94\linewidth]{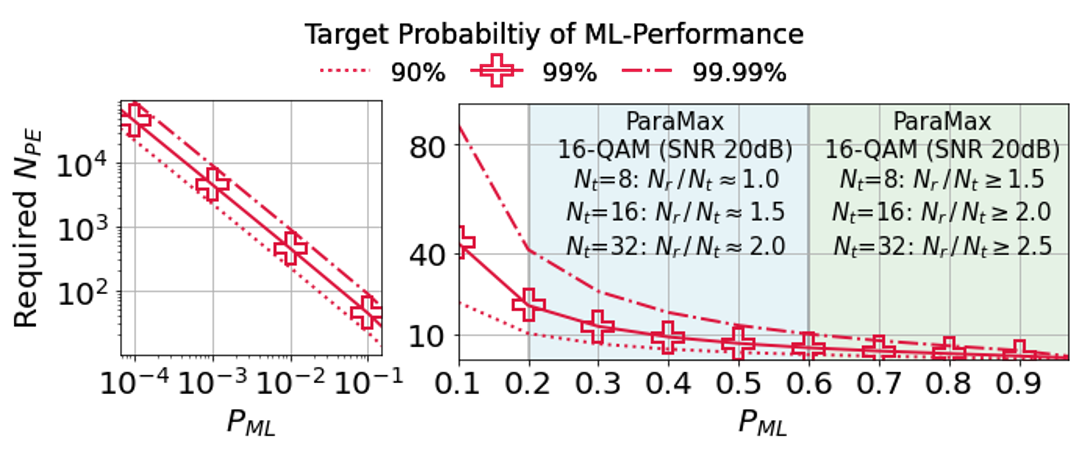}
    \caption{\normalfont Required $N_{PE}$ for \systemname{} to perform near-ML detection.}
    \label{f:required_npe_for_ml}
    \end{subfigure}
    \caption{\normalfont \textbf{Detection Sampling Evaluation.} Figure~\ref{f:prob_ml_comparison} plots $P_\text{ML}$ for three heuristic-based detectors (colors) varying modulations (hatch patterns), SNRs, and $N_V$. Figure~\ref{f:prob_ml_varying_ratio} shows impact of $N_r/N_t$ (from 1 to 4, from Large to Massive MIMO) on ParaMax's $P_\text{ML}$ for different $N_t$ user numbers (colors) and SNRs (hatch patterns).
    Figure~\ref{f:required_npe_for_ml} plots relationship between $P_\text{ML}$ and $N_{PE}$ for ML performance with various $\mathcal{P}_\mathbf{T}({\text{ParaMax}_\text{ML}})$. Approximate MIMO feasibility of \systemname{} is provided for two blocks of  high $P_\text{ML}$.} 
\label{f:eval_sampling}
\end{figure}

\parabreak{} In Figure~\ref{f:required_npe_for_ml},\footnote{Note that the formulas hold also for any non-parallel iterative method with independent sampling, where $N_{PE}$ is simply the number of required repetitions.} we plot $P_\text{ML}$ and corresponding required $N_{PE}$ for different $\mathcal{P}_\mathbf{T}({\text{ParaMax}_\text{ML}})$.


\parahead{Very Large MIMO with Low-Order Modulations.}
Figure~\ref{f:prob_ml_comparison} plots $P_\text{ML}$ as a function of $N\times N$ Large MIMO detection with different heuristic\hyp{}based detectors (SA, \systemname{}, and 2R\hyp{}\systemname{}) for various $N_V$ and modulations. Surprisingly, for the BPSK and QPSK modulations, all tested heuristic detectors achieve $P_\text{ML} \approx 1.0$, which implies nearly all PMIS runs successfully reach the ML-solution. For \systemname{} and 2R\hyp{}\systemname{}, this tendency is observed up to $N_V =512$ while we plot here only up to $N_V=128$ to save space. Only a few processing elements are enough to perform ML-detection up to $512 \times 512$ MIMO with BPSK and $256 \times 256$ MIMO with QPSK. While 
\systemname{} becomes currently unpractical at around $N_V=160$ (Figure~\ref{f:compute_vs_nv}), this MIMO size and its requirement for optimal detection is promising for city-scale \emph{Internet of Things} (IoT) applications envisioned in 5G networks or beyond. Those scenarios will handle hundreds or thousands of devices per BS with low-order modulations~\cite{miller2015internet,mazaheri2019millimeter,miorandi2012internet}, and may accept longer processing time than ordinary data communications.

\parahead{From Large to Massive MIMO with 16-QAM.} In the case of 16-QAM in Figure~\ref{f:prob_ml_comparison}, $P_\text{ML}$ notably drops as $N_V$ increases for all heuristic-based detectors and we observe higher $P_\text{ML}$ for 2R\hyp{}\systemname{}, \systemname{}, and SA.
Given that Large MIMO detection with high-order modulations is a challenging problem in general, we add more receiver antennas ($N_r$) to see the impact of $N_r$/$N_t$ ratio on $P_\text{ML}$. Figure~\ref{f:prob_ml_varying_ratio} shows this relationship for various user numbers for different SNRs. As $N_r/N_t$ increases (\emph{i.e,} from Large MIMO to Massive MIMO), $P_\text{ML}$ rapidly increases and then is converged to 1.0. While $P_\text{ML}$ for larger number of users ($N_t$) at lower SNRs tends to increase slower, $P_\text{ML} \approx 10^{-2}$ can still be achieved around $N_r/N_t = 2$, where the required $N_{PE}$ for ML-detection is around 1,000 (see Figure~\ref{f:required_npe_for_ml}, where we summarize the applicability of \systemname{}). We observed that the trace-driven channel with noise shows better performance (faster convergence than 20~dB SNR). 

\vspace{-0.3cm}

\begin{figure}
\centering
    \qquad
    \begin{subfigure}[b]{\linewidth}
    \centering
    \includegraphics[width=0.95\linewidth]{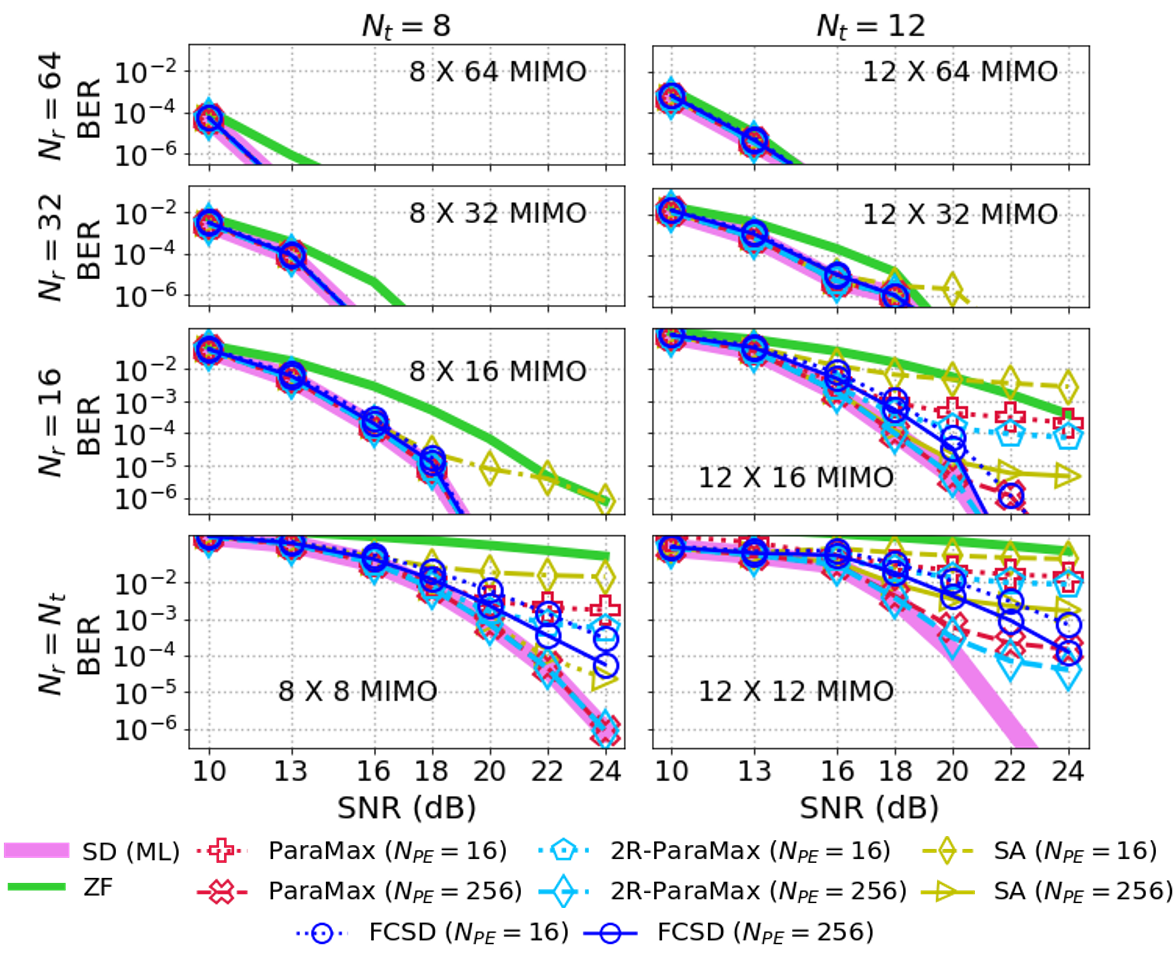}

    \caption{\normalfont 8-user and 12-user MIMO: BER as a function SNRs.}
    \label{f:massive_ber_snr}
    \end{subfigure}
    \qquad
    \begin{subfigure}[b]{\linewidth}
    \centering
    \includegraphics[width=0.93\linewidth]{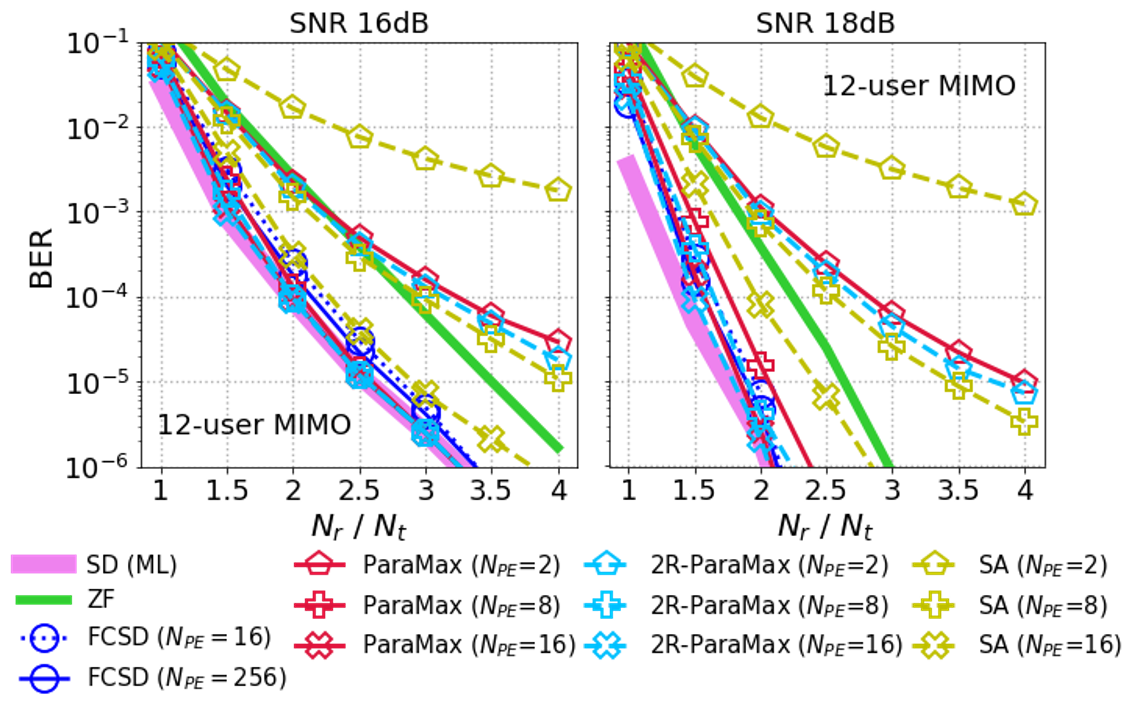}
    
    \caption{\normalfont 12-user MIMO: BER as a function of $N_r /N_t$ ratio.}
    \label{f:massive_ber_ratio}
    \end{subfigure}
\caption{\normalfont \textbf{(Overview) BER from Massive to Large MIMO.} Comparisons of detection BER in Large and Massive MIMO for various detectors across MIMO regimes and/or SNRs with 16-QAM.}
\label{f:ber}
\end{figure}

\subsection{Bit Error Rate (BER) Performance}
\label{s:eval_ber_throughput}

This section presents \systemnames{} detection BER. 
Recall that $N_{PE}$ is the number of processing elements (PEs) assigned to \systemname{} per subcarrier, where each PE performs a PMIS run. Since we assume fully-parallel \systemname{} for minimum detection latency, $N_{PE}$ is also equal to the number of PMIS runs. Note that regardless of computing platforms (CPU, GPU, or FPGA), the detection performance (BER and throughput) as a function of $N_{PE}$ is the same, as long as they can satisfy limited time requirements supporting all subcarriers (unless there exists a serious precision issue), while the definition of a single PE and total available PEs per device can vary depending on platforms and/or implementation details.
In the next subsection (Sec~\ref{s:eval_throughput}), we evaluate \systemname{} on multi-subcarrier systems, considering its detection latency, available parallelism, available compute time in wireless standards, and impact of
forward error control (FEC).   


\begin{figure}
    \centering
        \qquad
    
    
    \begin{subfigure}[b]{\linewidth}
    \includegraphics[width=0.86\linewidth]{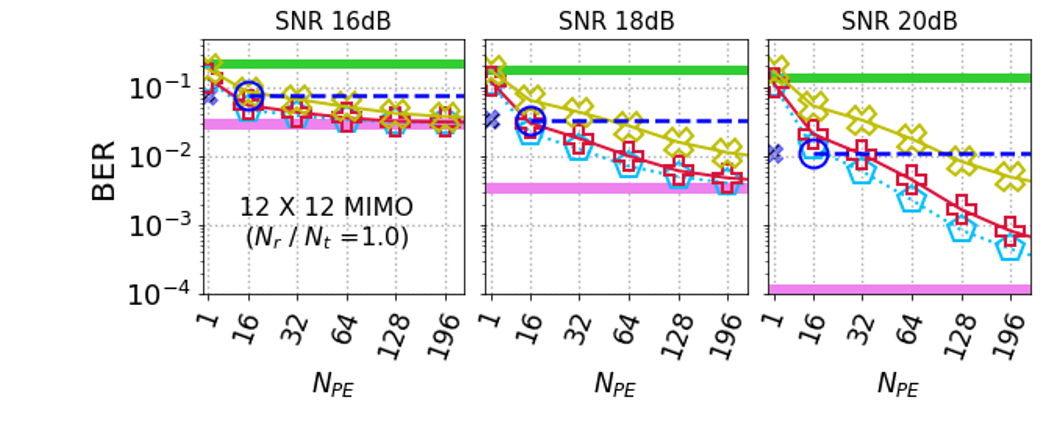}
    \caption{\normalfont 12-user Large MIMO: BER as a function of $N_{PE}$ varying SNRs.}
    \label{f:large_mimo}
    \end{subfigure}

    \begin{subfigure}[b]{\linewidth}
    \includegraphics[width=0.90\linewidth]{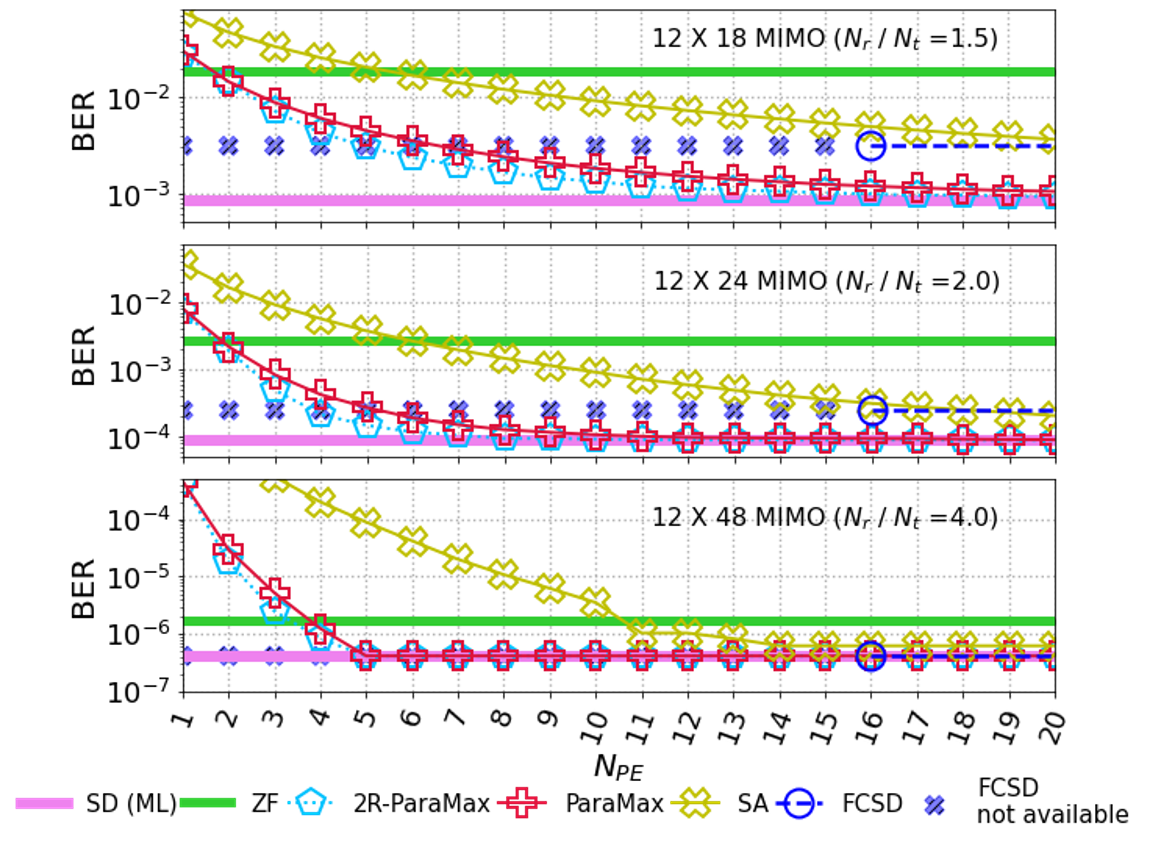}
    \caption{\normalfont 12-user Massive MIMO: BER as a function of $N_{PE}$ at SNR 16~dB.}
    \label{f:massive_ber_npe}
    \end{subfigure}
    
    \begin{subfigure}[b]{0.89\linewidth}
    \includegraphics[width=\linewidth]{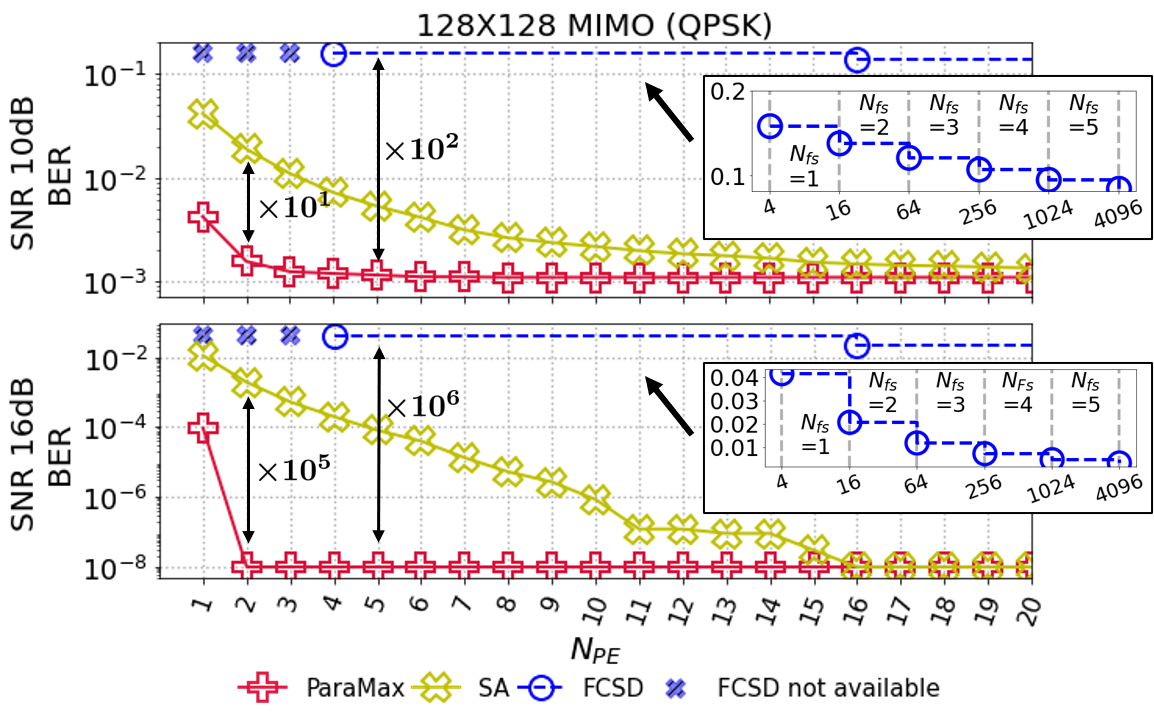}
    \caption{\normalfont 128 $\times$ 128 Very Large MIMO with QPSK modulation.}
    \label{f:very_large}
    \end{subfigure}
    \caption{\normalfont \textbf{(Detailed View) BER as a function of $N_{PE}$.} Comparisons of measured BER for various MIMO regimes and detectors with 16-QAM (Fig.~\ref{f:large_mimo},\ref{f:massive_ber_npe}) and QPSK (Fig.~\ref{f:very_large})  modulation.}
\label{f:detail_view}
\end{figure}




\parahead{Overview: BER from Massive to Large MIMO.} Figure~\ref{f:ber} shows BER performance in various MIMO regimes with 16-QAM. We consider \systemname{} and 2R-\systemname{} with 16 and 256 PEs, comparing them against other detectors such as ZF (linear), SA (heuristic), FCSD (tree search-based), and optimal SD (ML), where SA and FCSD (with channel ordering~\cite{dai2010simplified}) are comparison schemes of parallel architecture-based detectors. As expected, linear-based ZF, which requires high $N_r/N_t$ ratio for proper detection, performs poorly as the regime goes from Massive MIMO to Large MIMO (\emph{upper} to \emph{lower} in Figure~\ref{f:massive_ber_snr}) and with more users (\emph{left} to \emph{right} in Figure~\ref{f:massive_ber_snr}), showing several orders of magnitude worse BER performance against the other detectors, particularly at high SNRs.
In the case of the parallel architecture-based detectors, it is observed that for the same PEs, \systemname{} and 2R-\systemname{} outperform SA and FCSD detectors in all MIMO regimes and SNRs tested except that
in $12\times 12$ MIMO at high SNRs, FCSD outperforms \systemname{} and 2R-\systemname{} when with 16 PEs. However, 2R-\systemname{} reaches lower BER than FCSD when with 256 PEs. Figure~\ref{f:massive_ber_ratio} plots BER with 16-QAM as a function $N_r/N_t$ ratio with smaller $N_{PE}$ such as 2, 8, and 16. For low $N_r/N_t$ ratios, parallel architecture-based detectors even with 2~PEs can obtain lower BER than ZF. As the ratio increases, all detectors achieve better BER for the same $N_{PE}$, but more PEs are required to beat ZF.

\parahead{Detailed View: BER as a function of $N_{PE}$.} We evaluate BER as a function of $N_{PE}$ for 12-user Large and Massive MIMO in Figure~\ref{f:detail_view} to show the detailed performance comparison. Note that ZF is not suitable for parallelization, so it achieves the same performance, regardless of the number of PEs.
Figure~\ref{f:large_mimo} presents BER for $12 \times 12$ Large MIMO ($N_r=N_t$) with 16-QAM at various SNRs.
\systemname{} can support any number of PEs and approach the optimal performance as $N_{PE}$ increases (\emph{i.e.,} fine parallelism granularity), while the FCSD requires at least 16~PEs to operate the fully-parallel algorithm for the minimum $N_{fs}$, and the FCSD with $N_{PE}$=$16^{1}$ performs equivalently until $N_{PE}$ reaches $16^2$ (\emph{i.e.,} no gain between 16~PEs and 256~PEs). Figure~\ref{f:massive_ber_npe} focuses on Massive MIMO ($N_r\geq N_t$), showing the impact of $N_r/N_t$ ratio on both BER and $N_{PE}$. Higher ratios (\emph{upper} to \emph{lower} in Figure~\ref{f:massive_ber_npe}) lead to lower BER for the same PEs and smaller $N_{PE}$ for the near-ML BER, especially compared against $12\times 12$ Large MIMO (Figure~\ref{f:large_mimo}). Precisely, to reach the near-ML BER at SNR 16~dB for 12 users, 12-BS antenna MIMO requires around 60~PEs, 18-BS antenna MIMO requires 18~PEs, and 48-BS antenna MIMO requires only 5~PEs. Compared to SA and FCSD, \systemnames{} BER drops more rapidly as $N_{PE}$ increases for any $N_r/N_t$ ratios. For example, to reach $\text{BER} \approx 2\cdot10^{-4}$ at $12\times 24$ MIMO, where the FCSD and SA requires (over) 16~PEs, \systemname{} requires 6~PEs.

We also test 128-user Very Large MIMO with the QPSK modulation in Figure~\ref{f:very_large}. As analyzed in Figure~\ref{f:prob_ml_comparison}, we observe that very small $N_{PE}$ can result in BER convergence for the QPSK, which is very likely the optimal BER, although we cannot evaluate SD because of extremely high complexity. At SNR 16~dB, \systemname{} achieves over five orders of magnitude better BER than SA at 2~PEs and over six orders of magnitude better than FCSD at 4~PEs.
Note that the FCSD even with thousands of PEs (\emph{i.e.,} with high $N_{fs}$) cannot reach the \systemnames{} performance with a single PE.   
\vspace{-0.3cm}

\begin{figure}
    \centering
    \begin{subfigure}[b]{0.91\linewidth}
    \centering
    \includegraphics[width=0.99\linewidth]{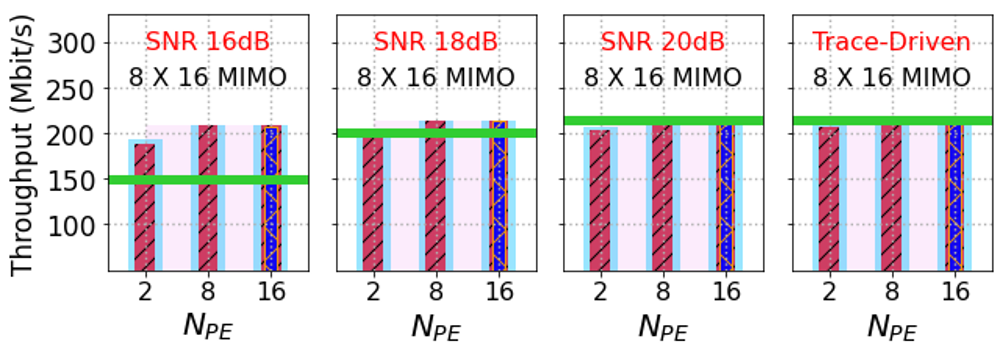}
    
    \caption{\normalfont Varying SNRs for 8 $\times$ 16 MIMO.}
    \label{f:throughput_comparison1}
    \end{subfigure}
    \qquad
    \begin{subfigure}[b]{0.90\linewidth}
    \includegraphics[width=\linewidth]{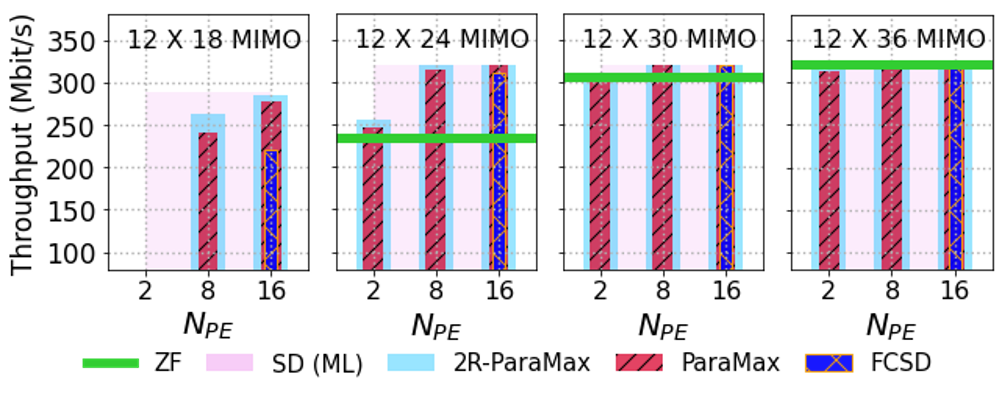}
    \caption{\normalfont Varying $N_r$ for 12-user MIMO at SNR 16dB.}
    \label{f:throughput_comparison2}
    \end{subfigure}
    
    \begin{subfigure}[b]{0.93\linewidth}
    \centering
    \includegraphics[width=0.92\linewidth]{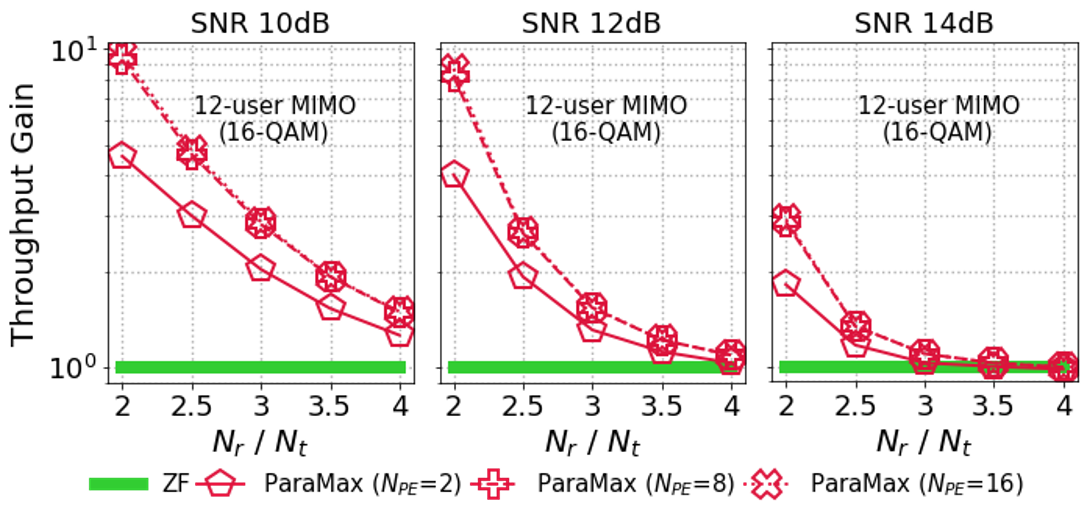}
    \caption{\normalfont  Achievable gain  (vs ZF) for the same number of subcarriers ($N_{SC}$).}
    \label{f:thr_gain}
    \end{subfigure}
    
    \caption{\normalfont \textbf{System Throughput of Massive MIMO in WLAN.} Achievable system throughput comparisons of various detectors as a function of $N_{PE}$ with 16-QAM in different scenarios (varying MIMO sizes or SNRs) for minimum detection latency.}

\label{f:throughput_comparison}
\end{figure}

\subsection{System Throughput Performance}
\label{s:eval_throughput}

This section evaluates throughput on multi-subcarrier systems. While detection BER is a fundamental metric for MIMO detection, the detector of the lowest BER does not necessarily imply the best throughput scheme, since real-world wireless systems include FEC techniques for error correction at the channel decoder under MIMO detector. 
Further, since the systems support many subcarriers with limited compute time, the total required computing resources to support them are another important metric for evaluation.

\begin{figure}
    \centering

    \qquad
    \begin{subfigure}[b]{0.92\linewidth}
    \includegraphics[width=\linewidth]{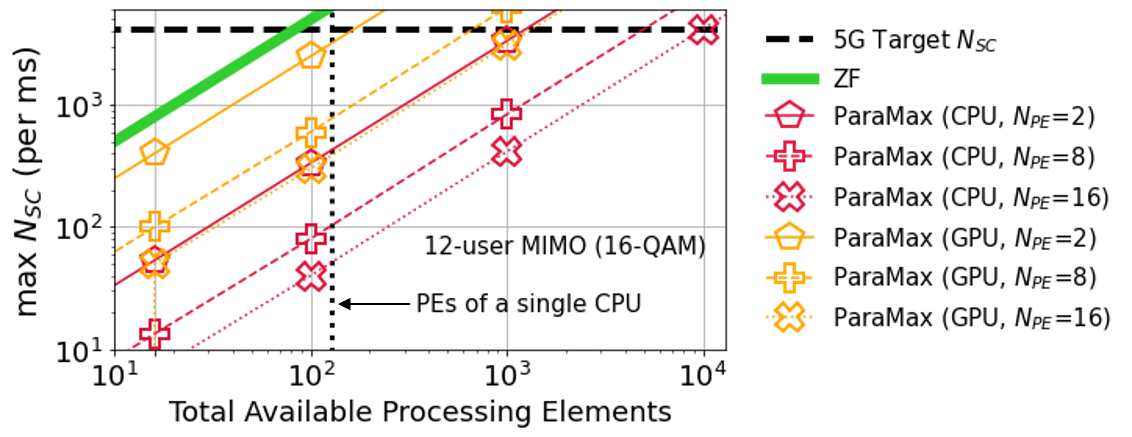}
    \caption{\normalfont Maximum number of subcarriers ($N_{SC}$) supported per millisecond.}
    \label{f:max_subcarrier}
    \end{subfigure}
    
    \begin{subfigure}[b]{\linewidth}
    \centering
    \includegraphics[width=0.95\linewidth]{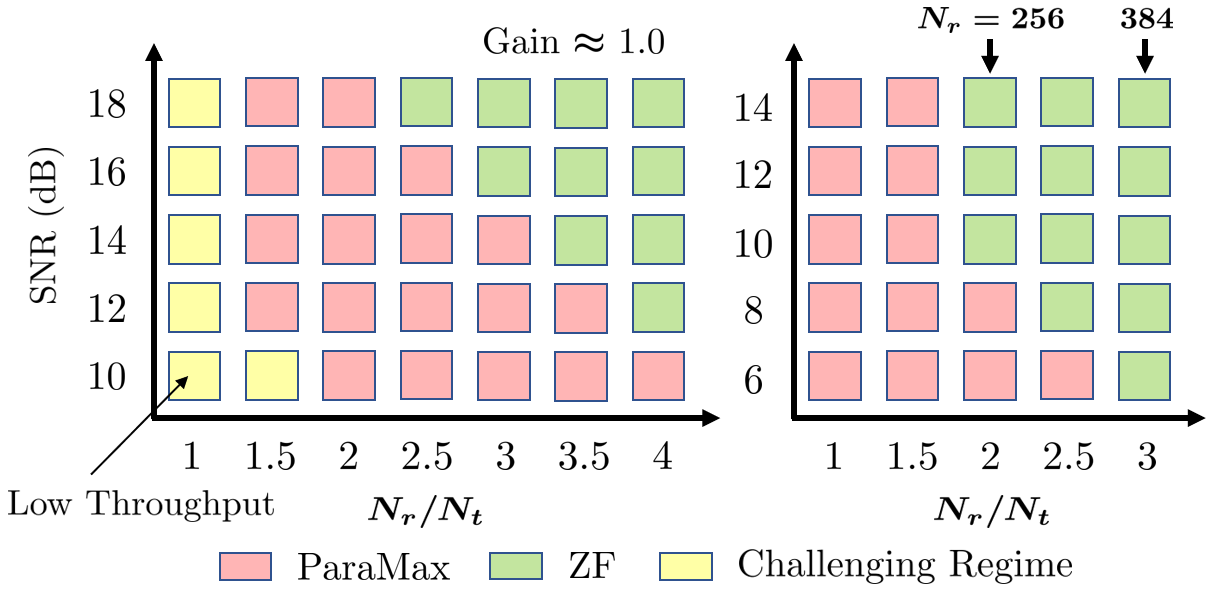}
    \caption{\normalfont Best throughput scheme for various MIMO/SNR regimes: 12-user MIMO with 16-QAM (left) and 128-user MIMO QPSK (right).}
    \label{f:best_16}
    \end{subfigure}


    \caption{\normalfont \textbf{Projection to 5G Systems.}{ Fig~\ref{f:max_subcarrier} shows maximum number of subcarriers that \systemname{} can support as a function of total available PEs on computing devices. Fig~\ref{f:best_16} presents the estimated best\hyp{}throughput detectors for various regimes, supporting 5G target $N_{SC}$, on a CPU platform with ten state-of-the-art 128-core CPUs ($\text{total PEs}\approx 10^3$). Here, \systemnames{} $N_{PE}\leq4$}.}
\label{f:5G}
\end{figure}

We first consider a WLAN wireless system with 64 OFDM subcarriers with 1/2 rate convolutional coding, where optimal achievable (ML-based) throughput on the system has been measured via over-the-air experiments in \cite{flexcore-nsdi17}. We translate the measured detection BER into the corresponding convolutional code-applied BER (\emph{i.e.,} coded BER). Among the provided data we select achievable optimal throughput for $8 \times 8$ MIMO and $12 \times 12$ MIMO at SNR 21.6~dB as a baseline throughput, assuming the coded BER of optimal Sphere Decoder (SD) we test at the same scenario (\emph{i.e.,} same MIMO size and SNR) is close to their optimal coded BER. We compute the achievable optimal throughput for various scenarios, considering SNR and optimal \emph{Frame Error Rate} (FER) difference (ratio) against the baseline, for the frame size of 1500-byte and FER obtained from our coded BER. Then we compute throughput of various detectors considering FER difference between SD and corresponding detectors at each scenario. In WLAN scenarios, we maintain the minimum detection latency. For this, the system is expected to have total PEs of $N_{PE} \times (N_{SC}=64)$ on a computing platform, where $N_{SC}$ is the number of subcarriers. For example, for \systemname{} to support 2 PMIS runs (\emph{i.e.,} 2~PEs) per subcarrier, a single state-of-the-art CPU (with 128 cores) is enough, while multiple CPUs are required to support more PEs per subcarrier. Since MIMO detection is completely independent across subcarriers, \systemnames{} job scheduling and allocation for multiple devices are quite straightforward.

Figure~\ref{f:throughput_comparison} shows throughput comparison against ZF and FCSD for various scenarios in Massive MIMO.  Figure~\ref{f:throughput_comparison1} demonstrates the impact of SNRs for the given MIMO size. While for \systemname{} and 2R-\systemname{}, 8 PEs (per subcarrier) are enough to reach the optimal performance for all tested SNRs, ZF could not reach the optimal performance except over SNR 20~dB including trace-driven channel and noise.  Figure~\ref{f:throughput_comparison2} shows the impact of $N_r/N_t$ ratio varying receiver antenna numbers $N_r$ at fixed SNR 16~dB. As seen in the previous section, less PEs are required at high $N_r/N_t$ ratios to achieve the near-ML performance. We observe that for the same scenarios, even less PEs are required to achieve the near-optimal ``throughput'' performance (Figure~\ref{f:throughput_comparison2}) than to achieve the near-optimal ``BER'' performance (Figure~\ref{f:massive_ber_npe}) because of the impact of FEC. Figure~\ref{f:thr_gain} plots \systemnames{} throughput gains versus ZF for various $N_r/N_t$ ratios and SNRs. The high gains are achieved at low SNRs and/or low $N_r/N_t$ ratios. 
These throughput gains can be generalized to any $N_{SC}$ (even at different standards with assumptions/modifications), as long as both schemes can support all subcarriers satisfying the corresponding limited time requirements.

In general, cellular\hyp{}networked systems, such as 4G or 5G systems, support many more subcarriers, allowing more compute time than WLAN. To project the throughput performance onto 5G scenarios, we plot the maximum $N_{SC}$ that can be supported per millisecond with 5G target, as a function of total available PEs on a computing platform, considering detector's latency and parallelism based on assigned PEs per subcarrier in Figure~\ref{f:max_subcarrier} (we report GPU-\systemname{} based on extrapolated data, as discussed in section \ref{s:eval_latency}). The figure implies how many computing resources are required to support 5G systems. A ZF-based system can support 5G target $N_{SC}$ even with a single state-of-the-art 128-core CPU due to its short latency and minimum PE usage. In the case of \systemname{}, tens of CPUs are required to support 2 PEs per subcarrier and hundreds for 16 PEs.

Furthermore, we estimate the best throughput scheme (ZF vs. \systemname{}) for various MIMO regimes and SNRs, assuming a computing platform with ten CPUs in Figure~\ref{f:best_16} (\emph{left}) for 12-user 16-QAM and Figure~\ref{f:best_16} (\emph{right}) for 128-user QPSK based on achievable \systemname{} throughput gains (vs ZF), although 128-user QPSK MIMO is currently unpractical (Figure~\ref{f:compute_vs_nv}).  
For gain $\approx$ 1.0, we report the ZF as the best scheme since it takes less compute time, while we report challenging regimes where ZF does not perform well and \systemname{} requires at least several tens of PEs per subcarrier for the near-ML performance. We observe that \systemname{} enables many challenging regimes of ZF (\emph{i.e.}, low $Nr/Nt$ ratio and/or low SNRs) by assigning reasonably more PEs.\footnote{Advanced FEC schemes such as LDPC and Polar codes that are applied in 5G systems can enable the near-ML performance with ZF for more MIMO regimes and SNRs. However, even more users and lower SNRs will keep bringing out the same scenarios, where \systemname{} outperforms ZF, due to the fundamental detection BER gap.} In the case of the QPSK, at $Nr/Nt=2$ (relatively small ratio), ZF can outperform \systemname{} at some SNRs, but for this, 256-BS antennas are required to support 128 users, which is the double size $N_r$ of the-state-of-the arts. Of course, \systemname{} requires more computing resources (10\hyp{}100$\times$) than ZF, but the trend at emerging system-on-chip architectures with more and more PEs, as well as C-RAN architectures promisingly envisioned in 5G, support the direction of massively parallel architectures-based designs requiring low interaction among PEs. 

\vspace{-0.1cm}

\section{Discussion}
\label{s:discuss}

In this section, we investigate several challenges and opportunities of \systemname{} that are likely to further advance the system.

\parahead{Fully-optimized and adaptive ParaMax.} 
Considering that parallel tempering-related parameters are selected within a challenging scenario (16-QAM) in Section~\ref{s:PMIS}, \systemname{} could be fully\hyp{}optimized  for many different scenarios based on given user numbers, $N_r/N_t$ ratios, SNRs, modulation sizes, available total PEs, and/or wireless standards.
Moreover there is a well known trade-off between number of sweeps (latency) and required $N_{PE}$ for near-ML performance (compute resources) that could be explored (\emph{e.g.,} for 5G URLLC).

\parahead{Higher-order modulations.}
As modulation size increases, \systemnames{} detection is degraded rapidly and becomes not operable for Large MIMO with high-order modulations such as 64-QAM or higher, requiring over $10^{3}$ PEs even for $4\times4$ Large MIMO to achieve near ML-performance. For Massive MIMO, it is expected that even higher $N_r/N_t$ ratios than 16-QAM are required. Perhaps, more replicas or Metropolis sweeps ease the problem along with further optimization on \systemnames{} free-parameters related to parallel tempering such as temperature range for PMIS tuning. However, these gains will be obtained at the expense of longer latency.
An implementation of \systemname{} on dedicated hardware might improve the performance and reduce the computational cost order further.

\parahead{Compatibility with specialized hardware.} \systemname{} does not require any specific hardware. However, another important aspect of \systemname{} is that it is immediately compatible with future implementations that aim to deploy programmable specialized hardware (for Physics-based algorithms) designed to optimize problems in the Ising form
including quantum devices
such as quantum annealers~\cite{kim2019leveraging} and gate-model quantum computers running the QAOA algorithm~\cite{Hadfield:2017:QAO:3149526.3149530},
as well as novel paradigm of classical calculation such as Optical Coherent Ising Machines~\cite{hamerly2019}, CMOS-based annealers~\cite{aramon2019physics}, and Oscillator-based platforms~\cite{csaba2020coupled}. 



\section{Conclusion}
\label{s:conclusion}

In this work, we present \systemname{}, a soft MU-MIMO detector system for Large and Massive MIMO networks that first makes use of parallel tempering for MIMO detection. Our performance evaluation shows that \systemname{} enables currently-challenging MIMO regimes 
for commonly-used linear detectors, achieving the near-ML performance by assigning reasonably more compute resources. \systemname{} also outperforms conventional parallel architecture-based detectors such as FCSD and SA-based detectors, requiring less processing elements to achieve the near-ML performance.

\section*{Acknowledgements}
We thank the anonymous shepherd and reviewers of this paper, the NASA Quantum AI Laboratory (QuAIL), and the Princeton Advanced Wireless Systems (PAWS) Group for their extensive technical feedback. 
This research is supported by National Science Foundation (NSF) Award CNS\hyp{}1824357 and CNS\hyp{}1824470, and an award from the Princeton University School of Engineering and Applied Science Innovation Fund. Salvatore Mandr\'a is supported by NASA Ames Research Center (ARC), particularly the autonomous ATM part of the Transformative Tools and Technologies Project under the NASA Transformative Aeronautic Concepts Program, and DARPA under IAA 8839, Annex 125. Minsung Kim was also supported by the USRA Feynman Quantum Academy funded by the NAMS R\&D Student Program at NASA ARC (FA8750-19-3-6101).

\clearpage

\let\oldbibliography\thebibliography
\renewcommand{\thebibliography}[1]{%
  \oldbibliography{#1}%
  \setlength{\parskip}{0pt}%
  \setlength{\itemsep}{0pt}%
}

\begin{raggedright}
\bibliographystyle{concise2}
\bibliography{reference}

\begin{thebibliography}{10}
\expandafter\ifx\csname urlstyle\endcsname\relax
  \providecommand{\doi}[1]{doi:\discretionary{}{}{}#1}\else
  \providecommand{\doi}{doi:\discretionary{}{}{}\begingroup
  \urlstyle{rm}\Url}\fi

\bibitem{3gpp-ts-36-211}
{3GPP Technical Specification 36.211 version 11.5.0 Release 11: Evolved
  Universal Terrestrial Radio Access (E-UTRA) Physical channels and
  Modulation}.

\bibitem{abrao2010s}
T.~Abr{\~a}o, L.~D. de~Oliveira, F.~Ciriaco, B.~A. Ang{\'e}lico, P.~J.~E.
  Jeszensky, F.~J.~C. Palacio.
\newblock S/mimo mc-cdma heuristic multiuser detectors based on
  single-objective optimization.
\newblock \textit{Wireless personal communications}, \textbf{53}(4), 529--553,
  2010.

\bibitem{Agrell02}
E.~Agrell, T.~Eriksson, A.~Vardy, K.~Zeger.
\newblock Closest point search in lattices.
\newblock \textit{IEEE transactions on information theory}, \textbf{48}(8),
  2201--2214, 2002.

\bibitem{aramon2019physics}
M.~Aramon, G.~Rosenberg, E.~Valiante, T.~Miyazawa, H.~Tamura, H.~Katzgrabeer.
\newblock Physics-inspired optimization for quadratic unconstrained problems
  using a digital annealer.
\newblock \textit{Frontiers in Physics}, \textbf{7}, 48, 2019.

\bibitem{FCSD}
L.~Barbero, J.~Thompson.
\newblock Fixing the complexity of the sphere decoder for {MIMO} detection.
\newblock \textit{IEEE Transactions on Wireless Communications}, \textbf{7}(6),
  2131--2142, 2008.

\bibitem{barbero2008low}
L.~G. Barbero, T.~Ratnarajah, C.~Cowan.
\newblock A low-complexity soft-mimo detector based on the fixed-complexity
  sphere decoder.
\newblock \textit{2008 IEEE International Conference on Acoustics, Speech and
  Signal Processing}, 2669--2672. IEEE, 2008.

\bibitem{barbero2008extending}
L.~G. Barbero, J.~S. Thompson.
\newblock Extending a fixed-complexity sphere decoder to obtain likelihood
  information for turbo-mimo systems.
\newblock \textit{IEEE Transactions on Vehicular Technology}, \textbf{57}(5),
  2804--2814, 2008.

\bibitem{bjornson2015massive}
E.~Bj{\"o}rnson, E.~G. Larsson, M.~Debbah.
\newblock Massive mimo for maximal spectral efficiency: How many users and
  pilots should be allocated?
\newblock \textit{IEEE Transactions on Wireless Communications},
  \textbf{15}(2), 1293--1308, 2015.

\bibitem{botsinis2013quantum}
P.~Botsinis, S.~X. Ng, L.~Hanzo.
\newblock Quantum search algorithms, quantum wireless, and a low-complexity
  maximum likelihood iterative quantum multi-user detector design.
\newblock \textit{IEEE access}, \textbf{1}, 94--122, 2013.

\bibitem{boyarinov1981linear}
I.~Boyarinov, G.~Katsman.
\newblock Linear unequal error protection codes.
\newblock \textit{IEEE Transactions on Information Theory}, \textbf{27}(2),
  168--175, 1981.

\bibitem{chandra2001parallel}
R.~Chandra, L.~Dagum, D.~Kohr, R.~Menon, D.~Maydan, J.~McDonald.
\newblock \textit{Parallel programming in OpenMP}.
\newblock Morgan kaufmann, 2001.

\bibitem{crooks2007measuring}
G.~E. Crooks.
\newblock Measuring thermodynamic length.
\newblock \textit{Physical Review Letters}, \textbf{99}(10), 100,602, 2007.

\bibitem{csaba2020coupled}
G.~Csaba, W.~Porod.
\newblock Coupled oscillators for computing: A review and perspective.
\newblock \textit{Applied Physics Reviews}, \textbf{7}(1), 011,302, 2020.

\bibitem{dagum1998openmp}
L.~Dagum, R.~Menon.
\newblock Openmp: an industry standard api for shared-memory programming.
\newblock \textit{IEEE computational science and engineering}, \textbf{5}(1),
  46--55, 1998.

\bibitem{dahlman20134g}
E.~Dahlman, S.~Parkvall, J.~Skold.
\newblock \textit{4G: LTE/LTE-advanced for mobile broadband}.
\newblock Academic press, 2013.

\bibitem{dai2010simplified}
X.~Dai, S.~Cheung, T.~Yuk.
\newblock Simplified ordering for fixed-complexity sphere decoder.
\newblock \textit{Proceedings of the 6th International Wireless Communications
  and Mobile Computing Conference}, 804--808, 2010.

\bibitem{Damen03}
M.~O. Damen, H.~El~Gamal, G.~Caire.
\newblock On maximum-likelihood detection and the search for the closest
  lattice point.
\newblock \textit{IEEE Transactions on information theory}, \textbf{49}(10),
  2389--2402, 2003.

\bibitem{ding2020agora}
J.~Ding, R.~Doost-Mohammady, A.~Kalia, L.~Zhong.
\newblock Agora: Real-time massive mimo baseband processing in software.
\newblock \textit{Proceedings of the 16th International Conference on emerging
  Networking EXperiments and Technologies}, 232--244, 2020.

\bibitem{farhang2006markov}
B.~Farhang-Boroujeny, H.~Zhu, Z.~Shi.
\newblock Markov chain monte carlo algorithms for cdma and mimo communication
  systems.
\newblock \textit{IEEE transactions on Signal Processing}, \textbf{54}(5),
  1896--1909, 2006.

\bibitem{fincke1985improved}
U.~Fincke, M.~Pohst.
\newblock Improved methods for calculating vectors of short length in a
  lattice, including a complexity analysis.
\newblock \textit{Mathematics of computation}, \textbf{44}(170), 463--471,
  1985.

\bibitem{georgii2011gibbs}
H.-O. Georgii.
\newblock \textit{Gibbs measures and phase transitions}, vol.~9.
\newblock Walter de Gruyter, 2011.

\bibitem{guangda2009multi}
Y.~GuangDa, H.~FengYe, H.~JinFeng.
\newblock The multi-user detection for the mimo-ofdm system based on the
  genetic simulated annealing algorithm.
\newblock \textit{Proceedings. The 2009 International Workshop on Information
  Security and Application (IWISA 2009)}, 334. Citeseer, 2009.

\bibitem{Guo06}
Z.~Guo, P.~Nilsson.
\newblock Algorithm and implementation of the {K}-best sphere decoding for
  {MIMO} detection.
\newblock \textit{IEEE Journal on Selected Areas in Communications},
  \textbf{24}(3), 491--503, 2006.

\bibitem{Hadfield:2017:QAO:3149526.3149530}
S.~Hadfield, Z.~Wang, E.~G. Rieffel, B.~O'Gorman, D.~Venturelli, R.~Biswas.
\newblock Quantum approximate optimization with hard and soft constraints.
\newblock \textit{Workshop on Post Moores Era Supercomputing (PMES)}, 2017.
\newblock \doi{10.1145/3149526.3149530}.

\bibitem{hamerly2019}
R.~Hamerly, T.~Inagaki, P.~L. McMahon, D.~Venturelli, A.~Marandi, T.~Onodera,
  E.~Ng, C.~Langrock, K.~Inaba, T.~Honjo, K.~Enbutsu, T.~Umeki, R.~Kasahara,
  S.~Utsunomiya, S.~Kako, K.-i. Kawarabayashi, R.~L. Byer, M.~M. Fejer,
  H.~Mabuchi, D.~Englund, E.~Rieffel, H.~Takesue, Y.~Yamamoto.
\newblock Experimental investigation of performance differences between
  coherent ising machines and a quantum annealer.
\newblock \textit{arXiv preprint arXiv:1805.05217}, 2018.

\bibitem{hansen2009near}
M.~Hansen, B.~Hassibi, A.~G. Dimakis, W.~Xu.
\newblock Near-optimal detection in mimo systems using gibbs sampling.
\newblock \textit{GLOBECOM 2009-2009 IEEE Global Telecommunications
  Conference}, 1--6. IEEE, 2009.

\bibitem{SDComp1}
B.~Hassibi, H.~Vikalo.
\newblock On the sphere-decoding algorithm i. expected complexity.
\newblock \textit{IEEE transactions on signal processing}, \textbf{53}(8),
  2806--2818, 2005.

\bibitem{hastings1970monte}
W.~K. Hastings.
\newblock Monte carlo sampling methods using markov chains and their
  applications.
\newblock \textit{Biometrika}, 1970.

\bibitem{hedstrom2017achieving}
J.~C. Hedstrom, C.~H. Yuen, R.-R. Chen, B.~Farhang-Boroujeny.
\newblock Achieving near map performance with an excited markov chain monte
  carlo mimo detector.
\newblock \textit{IEEE Transactions on Wireless Communications},
  \textbf{16}(12), 7718--7732, 2017.

\bibitem{horn1999robust}
U.~Horn, K.~Stuhlm{\"u}ller, M.~Link, B.~Girod.
\newblock Robust internet video transmission based on scalable coding and
  unequal error protection.
\newblock \textit{Signal Processing: Image Communication}, \textbf{15}(1-2),
  77--94, 1999.

\bibitem{flexcore-nsdi17}
C.~Husmann, G.~Georgis, K.~Nikitopoulos, K.~Jamieson.
\newblock Flexcore: Massively parallel and flexible processing for large {MIMO}
  access points.
\newblock \textit{14th {USENIX} Symposium on Networked Systems Design and
  Implementation ({NSDI} 17)}, 197--211, 2017.

\bibitem{ising1925beitrag}
E.~Ising.
\newblock Beitrag zur theorie des ferromagnetismus.
\newblock \textit{Zeitschrift f{\"u}r Physik}, \textbf{31}(1), 253--258, 1925.

\bibitem{jalden2009error}
J.~Jald{\'e}n, L.~G. Barbero, B.~Ottersten, J.~S. Thompson.
\newblock The error probability of the fixed-complexity sphere decoder.
\newblock \textit{IEEE Transactions on Signal Processing}, \textbf{57}(7),
  2711--2720, 2009.

\bibitem{josuttis2012c++}
N.~M. Josuttis.
\newblock \textit{The C++ standard library: a tutorial and reference}.
\newblock Addison-Wesley, 2012.

\bibitem{karimi2017boosting}
H.~Karimi, G.~Rosenberg.
\newblock Boosting quantum annealer performance via sample persistence.
\newblock \textit{Quantum Information Processing}, \textbf{16}(7), 166, 2017.

\bibitem{kasi2020towards}
S.~Kasi, K.~Jamieson.
\newblock Towards quantum belief propagation for ldpc decoding in wireless
  networks.
\newblock \textit{arXiv preprint arXiv:2007.11069}, 2020.

\bibitem{katzgraber2006feedback}
H.~G. Katzgraber, S.~Trebst, D.~A. Huse, M.~Troyer.
\newblock Feedback-optimized parallel tempering monte carlo.
\newblock \textit{Journal of Statistical Mechanics: Theory and Experiment},
  \textbf{2006}(03), P03,018, 2006.

\bibitem{kim2019leveraging}
M.~Kim, D.~Venturelli, K.~Jamieson.
\newblock Leveraging quantum annealing for large mimo processing in centralized
  radio access networks.
\newblock \textit{Proceedings of the ACM Special Interest Group on Data
  Communication}, 241--255. ACM, 2019.

\bibitem{kim2020towards}
---{}---.
\newblock Towards hybrid classical-quantum computation structures in
  wirelessly-networked systems.
\newblock \textit{Proceedings of the 19th ACM Workshop on Hot Topics in
  Networks}, 110--116, 2020.

\bibitem{larsson2008fixed}
E.~G. Larsson, J.~Jalden.
\newblock Fixed-complexity soft mimo detection via partial marginalization.
\newblock \textit{IEEE transactions on Signal Processing}, \textbf{56}(8),
  3397--3407, 2008.

\bibitem{larsson2002maximum}
E.~G. Larsson, P.~Stoica, J.~Li.
\newblock On maximum-likelihood detection and decoding for space-time coding
  systems.
\newblock \textit{IEEE Transactions on Signal Processing}, \textbf{50}(4),
  937--944, 2002.

\bibitem{li2008reduced}
Q.~Li, Z.~Wang.
\newblock Reduced complexity k-best sphere decoder design for mimo systems.
\newblock \textit{Circuits, Systems \& Signal Processing}, \textbf{27}(4),
  491--505, 2008.

\bibitem{lucas2014ising}
A.~Lucas.
\newblock Ising formulations of many np problems.
\newblock \textit{Frontiers in Physics}, \textbf{2}, 5, 2014.

\bibitem{malkowsky2017world}
S.~Malkowsky, J.~Vieira, L.~Liu, P.~Harris, K.~Nieman, N.~Kundargi, I.~C. Wong,
  F.~Tufvesson, V.~{\"O}wall, O.~Edfors.
\newblock The world’s first real-time testbed for massive mimo: Design,
  implementation, and validation.
\newblock \textit{IEEE Access}, \textbf{5}, 9073--9088, 2017.

\bibitem{mandra2018deceptive}
S.~Mandra, H.~G. Katzgraber.
\newblock A deceptive step towards quantum speedup detection.
\newblock \textit{Quantum Science and Technology}, \textbf{3}(4), 04LT01, 2018.

\bibitem{mandra2017exponentially}
S.~Mandra, Z.~Zhu, H.~G. Katzgraber.
\newblock Exponentially biased ground-state sampling of quantum annealing
  machines with transverse-field driving hamiltonians.
\newblock \textit{Physical review letters}, \textbf{118}(7), 070,502, 2017.

\bibitem{mandra2016strengths}
S.~Mandra, Z.~Zhu, W.~Wang, A.~Perdomo-Ortiz, H.~G. Katzgraber.
\newblock Strengths and weaknesses of weak-strong cluster problems: A detailed
  overview of state-of-the-art classical heuristics versus quantum approaches.
\newblock \textit{Physical Review A}, \textbf{94}(2), 022,337, 2016.

\bibitem{masnick1967linear}
B.~Masnick, J.~Wolf.
\newblock On linear unequal error protection codes.
\newblock \textit{IEEE Transactions on Information Theory}, \textbf{13}(4),
  600--607, 1967.

\bibitem{mazaheri2019millimeter}
M.~H. Mazaheri, S.~Ameli, A.~Abedi, O.~Abari.
\newblock A millimeter wave network for billions of things.
\newblock \textit{Proceedings of the ACM Special Interest Group on Data
  Communication}, 174--186, 2019.

\bibitem{metropolis1949monte}
N.~Metropolis, S.~Ulam.
\newblock The monte carlo method.
\newblock \textit{Journal of the American statistical association},
  \textbf{44}(247), 335--341, 1949.

\bibitem{miller2015internet}
M.~Miller.
\newblock \textit{The internet of things: How smart TVs, smart cars, smart
  homes, and smart cities are changing the world}.
\newblock Pearson Education, 2015.

\bibitem{miorandi2012internet}
D.~Miorandi, S.~Sicari, F.~De~Pellegrini, I.~Chlamtac.
\newblock Internet of things: Vision, applications and research challenges.
\newblock \textit{Ad hoc networks}, \textbf{10}(7), 1497--1516, 2012.

\bibitem{mondal2009design}
S.~Mondal, A.~Eltawil, C.-A. Shen, K.~N. Salama.
\newblock Design and implementation of a sort-free k-best sphere decoder.
\newblock \textit{IEEE Transactions on Very Large Scale Integration (VLSI)
  Systems}, \textbf{18}(10), 1497--1501, 2009.

\bibitem{nee2000ofdm}
R.~v. Nee, R.~Prasad.
\newblock \textit{OFDM for wireless multimedia communications}.
\newblock Artech House, Inc., 2000.

\bibitem{nikitopoulos2018massively}
K.~Nikitopoulos, G.~Georgis, C.~Jayawardena, D.~Chatzipanagiotis, R.~Tafazolli.
\newblock Massively parallel tree search for high-dimensional sphere decoders.
\newblock \textit{IEEE Transactions on Parallel and Distributed Systems},
  \textbf{30}(10), 2309--2325, 2018.

\bibitem{plank2013screaming}
J.~S. Plank, K.~M. Greenan, E.~L. Miller.
\newblock Screaming fast galois field arithmetic using intel simd instructions.
\newblock \textit{FAST}, 299--306, 2013.

\bibitem{ramiro2015mimopack}
C.~Ramiro, A.~M. Vidal, A.~Gonzalez.
\newblock Mimopack: a high-performance computing library for mimo communication
  systems.
\newblock \textit{The Journal of Supercomputing}, \textbf{71}(2), 751--760,
  2015.

\bibitem{roger2012efficient}
S.~Roger, C.~Ramiro, A.~Gonzalez, V.~Almenar, A.~M. Vidal.
\newblock An efficient gpu implementation of fixed-complexity sphere decoders
  for mimo wireless systems.
\newblock \textit{Integrated Computer-Aided Engineering}, \textbf{19}(4),
  341--350, 2012.

\bibitem{rusek2012scaling}
F.~Rusek, D.~Persson, B.~K. Lau, E.~G. Larsson, T.~L. Marzetta, O.~Edfors,
  F.~Tufvesson.
\newblock Scaling up mimo: Opportunities and challenges with very large arrays.
\newblock \textit{IEEE signal processing magazine}, \textbf{30}(1), 40--60,
  2012.

\bibitem{sanders2010cuda}
J.~Sanders, E.~Kandrot.
\newblock \textit{CUDA by example: an introduction to general-purpose GPU
  programming}.
\newblock Addison-Wesley Professional, 2010.

\bibitem{samsung}
D.~Schoolar.
\newblock Massive mimo comes of age.
\newblock \textit{Samsung Official Whitepaper}, 2017.

\bibitem{argos-asilomar17}
C.~Shepard, J.~Ding, R.~Guerra, L.~Zhong.
\newblock Understanding real many-antenna {MU-MIMO} channels.
\newblock \textit{Proc. of the IEEE Asilomar Conf.}, 2017.

\bibitem{Argos}
C.~Shepard, J.~Ding, R.~E. Guerra, L.~Zhong.
\newblock Understanding real many-antenna mu-mimo channels.
\newblock \textit{Signals, Systems and Computers, 2016 50th Asilomar Conference
  on}, 461--467. IEEE, 2016.

\bibitem{argos-mobicom12}
C.~Shepard, H.~Yu, N.~Anand, E.~Li, T.~Marzetta, R.~Yang, L.~Zhong.
\newblock Argos: Practical many-antenna base stations.
\newblock \textit{Proceedings of the 18th annual international conference on
  Mobile computing and networking}, 53--64. ACM, 2012.

\bibitem{studer2006soft}
C.~Studer, M.~Wenk, A.~Burg, H.~Bolcskei.
\newblock Soft-output sphere decoding: Performance and implementation aspects.
\newblock \textit{2006 Fortieth Asilomar Conference on Signals, Systems and
  Computers}, 2071--2076. IEEE, 2006.

\bibitem{vsvavc2013soft}
P.~{\v{S}}va{\v{c}}, F.~Meyer, E.~Riegler, F.~Hlawatsch.
\newblock Soft-heuristic detectors for large mimo systems.
\newblock \textit{IEEE Transactions on Signal Processing}, \textbf{61}(18),
  4573--4586, 2013.

\bibitem{swendsen1986replica}
R.~H. Swendsen, J.-S. Wang.
\newblock Replica monte carlo simulation of spin-glasses.
\newblock \textit{Physical review letters}, \textbf{57}(21), 2607, 1986.

\bibitem{trebst2006optimized}
S.~Trebst, M.~Troyer, U.~H. Hansmann.
\newblock Optimized parallel tempering simulations of proteins.
\newblock \textit{The Journal of chemical physics}, \textbf{124}(17), 174,903,
  2006.

\bibitem{vieira2014flexible}
J.~Vieira, S.~Malkowsky, K.~Nieman, Z.~Miers, N.~Kundargi, L.~Liu, I.~Wong,
  V.~{\"O}wall, O.~Edfors, F.~Tufvesson.
\newblock A flexible 100-antenna testbed for massive mimo.
\newblock \textit{2014 IEEE Globecom Workshops (GC Wkshps)}, 287--293. IEEE,
  2014.

\bibitem{SD}
E.~Viterbo, J.~Boutros.
\newblock A universal lattice code decoder for fading channels.
\newblock \textit{{{IEEE} Trans. Inf. Theory}}, \textbf{45}(5), 1639--1642,
  1999.

\bibitem{wang2004approaching}
R.~Wang, G.~B. Giannakis.
\newblock Approaching mimo channel capacity with reduced-complexity soft sphere
  decoding.
\newblock \textit{2004 IEEE Wireless Communications and Networking Conference
  (IEEE Cat. No. 04TH8733)}, vol.~3, 1620--1625. IEEE, 2004.

\bibitem{237878}
L.~{Wei}.
\newblock Coded modulation with unequal error protection.
\newblock \textit{IEEE Transactions on Communications}, \textbf{41}(10),
  1439--1449, 1993.

\bibitem{wei1993coded}
L.-F. Wei.
\newblock Coded modulation with unequal error protection.
\newblock \textit{IEEE Transactions on Communications}, \textbf{41}(10),
  1439--1449, 1993.

\bibitem{welsh1990computational}
D.~J. Welsh.
\newblock The computational complexity of some classical problems from
  statistical physics, 1990.

\bibitem{wolniansky1998v}
P.~W. Wolniansky, G.~J. Foschini, G.~D. Golden, R.~A. Valenzuela.
\newblock V-blast: An architecture for realizing very high data rates over the
  rich-scattering wireless channel.
\newblock \textit{1998 URSI international symposium on signals, systems, and
  electronics. Conference proceedings (Cat. No. 98EX167)}, 295--300. IEEE,
  1998.

\bibitem{BigStation}
Q.~Yang, X.~Li, H.~Yao, J.~Fang, K.~Tan, W.~Hu, J.~Zhang, Y.~Zhang.
\newblock {BigStation: Enabling} scalable real-time signal processing in large
  {MU-MIMO} systems.
\newblock \textit{ACM SIGCOMM Computer Communication Review}, \textbf{43}(4),
  399--410, 2013.

\bibitem{yang2017optimizing}
Z.-C. Yang, A.~Rahmani, A.~Shabani, H.~Neven, C.~Chamon.
\newblock Optimizing variational quantum algorithms using {Pontryagin's}
  minimum principle.
\newblock \textit{Physical Review X}, \textbf{7}(2), 021,027, 2017.

\bibitem{young2004absence}
A.~Young, H.~G. Katzgraber.
\newblock Absence of an almeida-thouless line in three-dimensional spin
  glasses.
\newblock \textit{Physical review letters}, \textbf{93}(20), 207,203, 2004.

\bibitem{zheng2017k}
B.~Zheng, M.~Wen, F.~Chen, N.~Huang, F.~Ji, H.~Yu.
\newblock The k-best sphere decoding for soft detection of generalized spatial
  modulation.
\newblock \textit{IEEE Transactions on Communications}, \textbf{65}(11),
  4803--4816, 2017.

\end{thebibliography}
\end{raggedright}
\appendix
\end{document}